\documentclass[12pt]{iopart}

\usepackage{iopams}
\usepackage{graphics}
\usepackage{graphicx}
\usepackage{dcolumn}
\usepackage{epsfig}
\usepackage{bm}
\usepackage{color}

\usepackage{amssymb}
\usepackage{lineno}
\usepackage{epsfig}
\usepackage{subfigure}
\begin{document}

\title[]{Predicted thermoelectric properties of Olivine type Fe$_2$Ge$Ch_4$ ($Ch$ = S, Se and Te)}

\author{Vijay Kumar Gudelli and V. Kanchana$^*$}

\address{Department of Physics, Indian Institute of Technology Hyderabad, Ordnance Factory Estate, Yeddumailaram-502 205, Telangana, India}

\author{G. Vaitheeswaran}
\address{Advanced Centre of Research in High Energy Materials (ACRHEM), University of Hyderabad, Prof. C. R. Rao Road, Gachibowli, Hyderabad - 500 046, Telangana, India}

\ead{kanchana@iith.ac.in}
\begin{abstract}
We present here the thermoelectric properties of olivine type Fe$_2$Ge$Ch_4$ ($Ch$ = S, Se and Te) using the linear augmented plane wave method based on first principles density functional calculations. The calculated transport properties using semi-local Boltzmann Transport equation reveal very high thermopower for both S and Se based compounds compared to Te counterpart. The reason for this high thermopower is mainly because of the quasi flat nature of the bands at the valence and conduction band edges. The calculated thermopower of Fe$_2$Ge$S_4$ is in good agreement with the experimental reports at room temperature with the carrier concentration around $10^{18}-10^{19} cm^{-3}$. All the investigated systems show anisotropic nature in their electrical conductivity resulting in a value lesser of the order of 10$^2$ along the a-axis compared to the $b$ and $c$-axes. Among the studied compounds, Fe$_2$Ge$S_4$ and Fe$_2$Ge$Se_4$ emerge as promising candidates with good thermoelectric performance.
\end{abstract}

\vspace{2pc}

\maketitle

\section{Introduction}
Olivine minerals are well known as magnetic semiconducting materials. The magnetic semiconducting nature of the compounds generally find applications in optoelectronic and magnetic devices\cite{Furdyna}. The general formula for the olivine structure is A$_2$BC$_4$, and crystallise mostly in orthorhombic structure, where $A$ is a transition metal, $B$ represents p-orbital elements and $Ch$ is the chalcogen. The arrangement of the atoms in the crystal is very closely packed and resembles hexagonal close packed arrangement. Among the olivine structures, $Fe$ based olivines are well known because of their diverse magnetic nature at low and high temperatures\cite{Tore1,Tore2,Quintero}. In the case of Fe$_2$GeS$_4$, it was found to possess weak ferromagnetic nature upto 69 $K$, anti-ferromagnetic nature between 69-143 $K$ and above this temperature it is reported to be in paramagnetic nature\cite{Junod}. A ferromagnetic curie temperature is around 149.9 $K$ in Fe$_2$GeTe$_4$\cite{Quintero}. A similar Fe-based olivine type silicate Fe$_2$SiO$_4$ is well known for its magnetic and optical properties\cite{Guo}. The popularly known structure of pyrite FeS$_2$ with sulphur vacancy has a close relation with the olivine structure, which was explained extensively through the theoretical and experimental study by Yu et al\cite{Yu}., and they also reported the application of this material as a photovoltaic absorber. Following the above mentioned work, experimental study showed that the nano-structured Fe$_2$GeS$_4$ can be potentially used as photovoltaic material\cite{Fredrick}. A similar study on highly crystalline nano-structured Fe$_2$GeS$_4$ was performed experimentally by Park and co-workers\cite{Park}. It is quite certain that there are interesting features that can be established in the olivine structure other than the well-known magnetic properties. Apart from the magnetic and the recent photovoltaic studies, there are no further studies available for these materials. We are interested in studying the thermoelectric properties of iron based olivine structures of Fe$_2$GeCh$_4$ (Ch=S, Se and Te). The present study focus mainly on the prediction of the thermoelectric properties, where the motivation stems from the experimental study which has shown Fe$_2$SiS$_4$ and Fe$_2$GeS$_4$ to possess good thermopower\cite{thesis}, enabling us to study the thermoelectric properties of the above mentioned materials.

	The performance of a thermoelectric (TE) material mainly depends on the figure of merit ZT, given by $ZT = S^2$ $\sigma T$ $/ \kappa$.  Here $S$, $\sigma$, $\kappa$ and T refers to thermopower, electrical conductivity, thermal conductivity, and absolute temperature, respectively. $\kappa$ includes both the electronic $\kappa_e$, and the lattice contributions $\kappa_l$, i.e., $\kappa=\kappa_e + \kappa_l$. For good thermoelectric materials, the typical value of $ZT$ is around $1$ and above. To achieve figure of merit close to unity or above, we need materials to meet the requirement of high thermopower around $200 \mu V/K$  and above, together with high electrical conductivity and low thermal conductivity. These are the challenges to the present researchers to search for different class of materials to achieve the conflicting requirements of high thermopower and electrical conductivity along with low thermal conductivity. The successful thermoelectric materials having figure of merit close to unity includes, Bi$_2$Te$_3$, TAGS-85 (Tellurium-antimony-germanium-silver)\cite{TAGS}, filled skutterudites\cite{Sales}, PbTe/PbSe\cite{YPei}, etc. To explore the thermoelectric properties of iron based olivine structures, we have used the first principle based electronic structure calculations using semi-classical Boltzmann transport equations. The rest of the paper is organised as follows: section 2  describes  the methodology, and section 3 presents the results and discussions. Conclusions are given in section 4.

\section{Computational details}
Total energy calculations based on first principle density functional theory (DFT) were performed using the full-potential linear augmented plane wave (FP-LAPW) method as implemented in WIEN2k\cite{Blaha}. The total energies are obtained by solving the Kohn-Sham equations self-consistently.
The self-consistent calculations included  spin-orbit coupling. Since the calculations using standard local-density (LDA) or generalised gradient approximation (GGA) schemes for the exchange-correlation potential underestimate the band gaps of semiconductors, we have used the Tran-Blaha modified Becke-Johnson\cite{Becke} potential (TB-mBJ)\cite{Tran1} on top of GGA-PBE(Perdew-Burke-Ernzerhof) \cite{Perdew}, which is quite good at reproducing the experimental band gaps. For k-space integrations, a $6\times11\times13$ k-mesh was used for Fe$_2$GeCh$_4$ in the Monkhorst-Pack scheme \cite{Monkhorst}, resulting in 168 k-points in the irreducible part of the Brillouin zones for all the compounds, respectively. All the calculations were performed with an energy convergence criterion of $10^{-6}$ Ry per formula unit. The carrier concentration (both holes and electrons) and temperature ($T$) dependent thermoelectric properties like thermopower (S), transport functions ($\frac{\sigma}{\tau}$) ($\sigma$ is the conductivity, and $\tau$ is an inverse scattering rate) were computed using BOLTZTRAP\cite{Madsen} code, within the Rigid Band Approximation (RBA)\cite{Scheidemantel,Jodin,Chaput} and the constant scattering time ($\tau$) approximation (CSTA).\cite{Khuong,singh,aggate2} More details about the approximations of RBA and CSTA can be found in Ref. \cite{FeX2}, and the references therein. The crystal structures are generated using the VESTA\cite{vesta} software and the charge density plots are generated with the help of Xcrysden molecular structure visualization program\cite{xcrysden}.

\section{Results and discussions}
\subsection{Crystal structure and electronic structure of Fe$_2$GeCh$_4$}
Olivine structure of Fe$_2$GeCh$_4$ is orthorhombic with space group $Pnma$. In this the cation $Fe$ and the anion $Ch$ forms an distorted octahedron Fe$Ch_6$. The $Fe$ in Fe$_2$GeCh$_4$ has two independent wyckoff positions, $4a$ and $4c$. The coordination number of these two sites are similar, but are found to be different in the bond length between $Fe$-$Ch$ and the orientation of the octahedron formed by them. 
The crystal structure of Fe$_2$GeS$_4$ is given in Fig. 1 showing the edge and vertex sharing octahedron. The bond length of $Fe-S$ formed by $Fe$ at $4a$-site varies from 2.477 to 2.551 $\AA$ which shows an average bond length of 2.494 $\AA$ and the distortion index of the octahedron is $0.011$. In case of $Fe$ at $4c$-site the variation of the bond length is from 2.449 to 2.629 with an average bond length of 2.539 $\AA$ and the distortion index is $0.024$. This is a clear evidence of the difference in the distortion of the octahedron formed by the two different $Fe$-sites. We also observed that the distortion of the $Fe$ at the $4a$-site is lesser than $4c$-site $Fe$, which is very similar to that of Fe$_2$SiS$_4$\cite{thesis}. The two different orientations of the octahedron formed by $Fe$ has a significant importance and differ in contribution to the bands at the Fermi level, which is discussed in detail in the later section. Earlier Yu et. al\cite{Yu}., explained about the close relation of olivine type Fe$_2$GeS$_4$ and Fe$_2$SiS$_4$ with the pyrite FeS$_2$, and also reported these two compounds to possess almost similar absorption coefficient as that of pyrite. We have found structural similarities between olivine and marcasite, which is a polymorphic phase of pyrite FeS$_2$. In both olivine and marcasite, we can see a distorted octahedron formed by $Fe$ with sulphur. The edge sharing octahedron of olivine along the $b$-axis is also seen in marcasite, see Fig. 1, together with the vertex sharing octahedron of olivine along the unit cell diagonal, which is also present in the similar direction of unit cell diagonal of the marcasite. This shows that the olivine type structures are having very close relation with both the polymorphic phases of marcasite and pyrite FeS$_2$. Our earlier work on the polymorphic phases of marcasite and pyrite has shown close relation between these two phases and also reported good thermoelectric properties for both the phases\cite{FeS2}.  This provoke us further to study the thermoelectric properties of olivine structures and compare the same with polymorphic phases of FeS$_2$.

\par
With the motivation as mentioned above, we intend to study the thermoelectric properties through the first principles electronic structure calculations. All the present calculations are performed at the experimental volume\cite{Fe2GeS4,Fe2GeSe4,Fe2GeTe4}. The electronic structure properties are carried out using the TB-mBJ exchange correlation functional. The band structure of all the compounds along the high symmetry directions are presented in Fig. 2. The corresponding energy gaps are found to be 2.01 $eV$ for Fe$_2$GeS$_4$, 1.69 $eV$ for Fe$_2$GeSe$_4$, and 0.6 $eV$ for Fe$_2$GeTe$_4$. The experimentally reported band gap for Fe$_2$GeS$_4$ was 1.40 $eV$ and the same group found it to be 1.36 $eV$ with the GGA+U method\cite{Yu}. In the present calculations, we found a slightly higher value with the TB-mBJ functional, which is also the case in other similar type of compounds with the same TB-mBJ functional\cite{Dixit}. The spin-orbit coupling (SOC) has shown a significant effect in Fe$_2$GeTe$_4$ compared to Fe$_2$GeS$_4$  and Fe$_2$GeSe$_4$. This resulted in a degeneracy lift in $Te-5p$ states about $0.24 eV$ in Fe$_2$GeTe$_4$.
The band structure of Fe$_2$GeS$_4$ and Fe$_2$GeSe$_4$ are almost similar to each other, whereas Fe$_2$GeTe$_4$ is found to be different. The valence band maximum (VBM) is at $Y$ and conduction band minimum (CBM) is observed along $\Gamma-Y$ for Fe$_2$GeS$_4$, which result in an indirect band gap semiconductor, in agreement with the earlier work\cite{Yu,thesis}. Similar behaviour of indirect band gap is also seen in Fe$_2$GeSe$_4$ along $\Gamma-Y$ of VBM and at $\Gamma$ in CBM. But Fe$_2$GeTe$_4$ is found to be a direct band gap semiconductor with the VBM and CBM both occurring at $\Gamma$ point. 
As indicated previously, the octahedron formed by $Fe$ and $Ch$ provides the hybridisation between $Fe-Ch$ states. According to the octahedronl crystal field splitting, the $Fe-3d$ orbitals are divided into three filled $t_{2g}$ and empty doublet $e_g$ states. The representation of the crystal field splitting of $Fe$-$3d$ is shown schematically in Fig. 3. The filled triplet states of $t_{2g}$ will contribute to the VBM as non-bonding states, with very small contribution from $Fe-3d_{e_g}$ and $S-3p$. The empty doublet states of $Fe-3d_{e_g}$ interact with the chalcogen-$p$ states and form the bonding states below the VBM. The higher energy states of $Fe-3d_{e_g}$ and chalcogen-$p$ forms the anti-bonding states which contribute to the CBM. The crystal field splitting is very similar to that of the prototype Rh$_2$ZnO$_4$, in the low spin state of the $Rh$\cite{Nagaraja,Volnianska}. 
The $Ge$-states reside much below in the valence band. 
\par
The corresponding density of states (DOS) of all the compounds is shown in Fig. 4. From this figure, it is evident that at both the band edges, $Fe-3d$ states are more dominant in all the compounds, but in the case of valence band, we also find the chalcogen-$p$ states to contribute a little (see Fig. 4(b)). The contribution of the $Fe$ and $S$ states at the VBM and CBM is almost very similar to that of the pyrite, since the $Ge$ character in Fe$_2$GeS$_4$ is much below the valence band. This shows the evidence of similarity between the olivine Fe$_2$GeS$_4$ and pyrite FeS$_2$\cite{Rosso,Eyret}.
To further analyse the contribution of each element to the total DOS, we have also plotted the orbital resolved DOS e.g. $m$-projected DOS of Fe$_2$GeS$_4$, which is shown in Fig. 4(c) and 4(d). 
The slight difference observed in the contribution of the two $Fe$ states in the DOS at the Fermi level may be due to the different orientation of the FeCh$_6$ octahedron resulting in different bond lengths of $Fe1$ and $Fe2$ with sulphur as mentioned above. The states below $-1 eV$ in the valence band are due to the $S-3p$ orbital.
The crystal orbital overlap populations (COOP) analysis of Fe$_2$SiS$_4$ has also shown similar type of bonding\cite{thesis}. We further investigated the nature of bonding among the elemental species in Fe$_2$GeS$_4$ and the charge density plot is shown in Fig. 5. This shows that the bonding between $Fe-S$ is a weak covalent bond, while $Se-Ge$ form strong covalent bond. The covalent bonding between $Fe-S$ is comparatively stronger along $b$ and $c$- axes of Fe$_2$GeS$_4$ (see Fig. 5(b) and 5(c)), which might indicate a better flow of charge carriers along these two axes compared to the other axis $a$. This might be due to the presence of edge-sharing octahedron along the $b$-axis which is layered through $c$-axis, resulting a strong covalent bonding between $Fe-S$ along the $b$- and $c$- axes, while along the $a$-axis the effect of edge and vertex sharing is weak which eventually lower the covalent bonding nature along this direction. In addition, it is to be noted that lattice parameter is larger along the $a$-axis leading to lesser interaction resulting in weak bond along the respective crystallographic direction, leading to a weak charge flow resulting in low electrical conductivity along $a$-axis which is discussed in the later section. From the band structure of Fe$_2$GeS$_4$ and Fe$_2$GeSe$_4$ it is quite evident that the dispersion of bands along the three crystallographic directions i. e. $\Gamma$-X, $\Gamma$-Y and $\Gamma$-Z are very flat in nature, which can result in a sharp increase in the DOS at the band edges. We observe almost an identical increase in DOS at both the edges in all the compounds, which might indicate a favourable condition for the band dependent properties like thermopower for both the carriers. 
\par The basis of the present study is to predict the thermoelectric properties of Fe$_2$GeCh$_4$ as a function of carrier concentration at various temperatures. For this purpose one need to study the effective mass of both the carriers at the band edges. We have calculated the effective mass at the conduction and valence band edges by fitting the energy of the respective bands to a quadratic polynomial in the reciprocal lattice vector $\vec{k}$. The calculated effective mass of the bands along $\Gamma-X$, $\Gamma-Y$ and $\Gamma-Z$ directions are shown in Table-I. Lower values of the effective mass are observed for CBM compared to VBM. The effective mass values of both VBM and CBM (except along $\Gamma-Z$ of CBM) are found to decrease from sulphur to tellurium among the investigated compounds along the three directions mentioned above. Effective mass of the carriers of both Fe$_2$GeS$_4$ and Fe$_2$GeSe$_4$ are almost similar (slightly lesser for Fe$_2$GeSe$_4$) which is because of the similar band structure of these two compounds, whereas in the case of Fe$_2$GeTe$_4$ we find the effective mass of the carriers to be lower, resulting from the highly dispersive bands along the specified directions as mentioned above. The higher values of the effective mass of the carriers at both band edges might indicate a high thermopower in Fe$_2$GeS$_4$ and Fe$_2$GeSe$_4$. With this preliminary idea, we have calculated the transport properties for optimised doping levels of the carriers and is discussed in the next section. The effective mass values of the olivine Fe$_2$GeS$_4$ are higher in comparison with the marcasite and pyrite FeS$_2$, which shows that the thermopower of olivine type of minerals may have higher value than the marcasite and pyrite, which is seen in the present calculation and the same is explained in the next section. 

\subsection{Thermoelectric properties of Fe$_2$GeCh$_4$}
We now look into the thermoelectric properties of Fe$_2$GeCh$_4$. 
The uniform increase in DOS at both the edges suggest that these materials might show favourable thermoelectric properties for both the carrier concentrations. This allow us to calculate the carrier concentration dependent thermoelectric properties such as thermopower ($S$ in $\mu$ $V/K$) and electrical conductivity scaled by relaxation time ($\sigma/\tau$ in $(\Omega m s)^{-1}$) using BoltzTrap code within the limit of RBA and CSTA as mentioned in section-II at various temperatures for both electrons and holes. As mentioned earlier, the olivine type structure is non-cubic, and it becomes very important to analyse the direction dependent thermoelectric properties e.g. along $a$, $b$ and $c$- directions. From the earlier studies it is quite evident that the direction dependent thermoelectric properties are very important due to the anisotropic nature of the systems\cite{SnSe,DJS_PRL,JAP_CuGaTe2}. For example, the delafossite type PtCoO$_2$ and PdCoO$_2$ revealed the importance of the anisotropic nature of the thermoelectric properties, where we can find a huge difference in the thermoelectric properties along the in-plane and out-of-plane directions\cite{DJS_PRL}. Keeping in view of the above consideration, we have calculated the direction dependent thermoelectric properties such as $S$ and $\sigma/\tau$, and are presented in Fig. 6-8, for all the compounds for both the holes (n$_h$ for holes) and electrons (n$_e$ for electrons) as carriers. From the thermopower of all the compounds, we observed that there is no bipolar conduction seen in the case of Fe$_2$GeS$_4$ and Fe$_2$GeSe$_4$ (see Fig. 6(a) and 7(a)) because of the higher band gaps ($>1 eV$), whereas in Fe$_2$GeTe$_4$ (see Fig. 8(a)) we see a bipolar conduction at higher concentration in the case of holes and at lower concentration in the case of electrons. The reason for the bipolar conduction in Fe$_2$GeTe$_4$ may be because of low band gap ($0.6 eV$). From Fig. 6, it is clear that the thermopower of Fe$_2$GeS$_4$ is found to vary from $850-250 \mu V/K$ for the optimum hole concentration of $10^{18}-10^{21} cm^{-3}$, whereas for electrons, it is found to be $800-200 \mu V/K$, and the range of the thermopower values are apparently very high at 300 $K$ and 500 $K$ temperatures in comparison with the traditional thermoelectric materials\cite{Sales,YPei,Stordeur,Testardi,Jeon,Plechcek}. For example the commercially used thermoelectric material Bi$_2$Te$_3$ has a thermopower of $225 \mu V/K$ at room temperature\cite{Tritt}, and we find the same value of the thermopower even at the high concentration around $10^{20} cm^{-3}$ in Fe$_2$GeS$_4$ at room temperature (the thermopower value is still higher at lower concentrations). 
This shows that Fe$_2$GeS$_4$ might be  having a very high thermopower for both the carriers. In the case of Fe$_2$GeSe$_4$, the thermopower is found to be slightly lower compared to Fe$_2$GeS$_4$ because of the effective mass being smaller as seen from Table-I, but it is also shown to be more promising for both the carriers. In the case of Fe$_2$GeTe$_4$, we found a lower thermopower values compared to the other two chalcogens, which is because of the lower values of the effective mass of the carriers in this compound. Thermopower is found to be below $550 \mu V/K$ at a hole concentration of $10^{18} cm^{-3}$, whereas we found slightly higher values compared to holes in the case of electrons. But there exist a bipolar conduction in the case of electrons at lower concentration region. Anisotropy in thermopower is seen in all the compounds along all three crystallographic axes. We also observed that the anisotropy is found to increase as we move down the chalcogens for both the carriers. 
\par
Earlier experiments on Fe$_2$GeS$_4$ have shown a bulk thermopower of $750 \mu V/K$ at room temperature and from our study also, we find the same value of thermopower at room temperature for the following concentrations of $1.14 \times 10^{19} cm^{-3}$ along $a$-axis, $9.31 \times 10^{18} cm^{-3}$ along $b$-axis and $5.15 \times 10^{18} cm^{-3}$ along $c$-axis for the holes\cite{thesis}. The above mentioned concentrations along the three directions can be achieved well within the experimental conditions for the semiconductors. In case of the electrons we found the carrier concentration to be below $10^{18} cm^{-3}$ for the same thermopower of $750 \mu V/K$ at room temperature. The calculated thermopower of $750 \mu V/K$ for Fe$_2$GeS$_4$ at room temperature is in good agreement with the measured single crystal bulk thermopower of $750 \mu V/K$ at a concentration around $5 \times 10^{18} cm^{-3}$\cite{Yu}. This shows that the predicted thermopower values are in line with the experimental results. In comparison  with the marcasite FeS$_2$ ($\sim 300 \mu V/K @ 500 K$), the thermopower of olivine type Fe$_2$GeS$_4$ ($\sim 460 \mu V/K @ 500 K$) has higher value at a hole concentration of $10^{20} cm^{-3}$. We find a similar value of the thermopower of pyrite FeS$_2$ ($\sim 500 \mu V/K @ 900 K$) with that of the olivine type Fe$_2$GeS$_4$ ($\sim 500 \mu V/K @ 900 K$) at the same concentration as mentioned above. This indicates that olivine can show better thermoelectric performance if it possess high electrical conductivity and low thermal conductivity. The electrical conductivity scaled by the relaxation time is discussed in the succeeding section.

\par
The electrical conductivity, in all the compounds is found to be more for electrons compared to the holes.
The difference in the electrical conductivity for the electrons is almost one order more compared to the holes. We also observe a huge anisotropy along $a$-axis compared to the $b$ and $c$-axes. The $\sigma/\tau$ along the $a$-axis is almost two orders lower throughout the optimum concentrations for both the carriers for all the compounds. This might due to the higher value of lattice parameter $'a'$ compared to $'b'$ and $'c'$ and also weak covalent bonding nature along this direction compared to other two axes as mentioned earlier. A similar situation of lower electrical conductivity along the direction of larger lattice parameter was also found in the case of $SnSe$ crystal\cite{SnSe}. As seen in the bonding, the interaction or the extent of hybridisation along the $a$-axis is also lesser. We found a significantly lower anisotropy between the $b$ and $c$-axes in all the compounds. This eventually confirm that the thermoelectric performance of the investigated compounds show better applications along the $b$ and $c$-axes compared to the $a$-axis. 
The value of electrical conductivity is found to increase as we move down the chalcogen group throughout the hole concentration. But in the case of electrons it is found to be a non-monotonic variation, where it decreases from Fe$_2$GeS$_4$ to Fe$_2$GeSe$_4$ followed by an increase in Fe$_2$GeTe$_4$ within the optimum electron concentration. Electrical conductivity of Fe$_2$GeTe$_4$ is found to be slightly higher compared to Fe$_2$GeS$_4$ in the case of electrons. But the bipolar nature of thermopower as seen in Fe$_2$GeTe$_4$ will not make it suitable for good thermoelectric performance. This allow us to state that among the investigated olivine type of structures both Fe$_2$GeS$_4$ and Fe$_2$GeSe$_4$ emerge as good thermoelectric candidates for both the charge carriers. In order to understand the lower values along the $a$-direction, we further estimated the order of the mobility scaled by the relaxation time ($\mu = R_H \times \sigma/\tau$) and is shown in Fig. 9. From this figure it is quite evident that the mobility of both carrier concentrations along the $a$-axis is very low compared to the other two axes, which resulted in two orders lower electrical conductivity along $a$-axis. The lower value of the mobility along the $a$-axis is consistent with the charge flow as mentioned in the section IIIA. 
\par
We further studied the power-factor ($S^2 \sigma/\tau$) of Fe$_2$GeS$_4$ and Fe$_2$GeSe$_4$ and is shown in Fig. 6(c) and 7(c). The power factor for electrons (above $10^{11} W/m K^{-2} s^{-1}$) is slightly higher compared to the holes (below $10^{11} W/m K^{-2} s^{-1}$) for Fe$_2$GeS$_4$, and it is because of the high electrical conductivity of electrons compared to the hole concentration of $10^{21} cm^{-3}$. But in the case of Fe$_2$GeSe$_4$ we find nearly equal values for both electrons as well as holes ($\sim 10^{11} W/m K^{-2} s^{-1}$). The calculated power factor of Fe$_2$GeS$_4$ and Fe$_2$GeSe$_4$ are found to be almost similar to the marcasite and pyrite structures of the FeS$_2$\cite{FeS2} and FeSe$_2$\cite{FeX2}. However, we have seen earlier that the thermopower of olivine type Fe$_2$GeS$_4$ has higher value compared to the marcasite and almost nearly equal value with pyrite FeS$_2$, but we find a lower values of power factor which is because of the lower electrical conductivity of Fe$_2$GeS$_4$ ($\sim 0.6\times 10^{17} (\Omega m s)^{-1} @ 500 K$ and $\sim 0.5\times 10^{17} (\Omega m s)^{-1} @ 900 K)$, compared to marcasite ($ \sim 1\times 10^{18} (\Omega m s)^{-1} @ 500 K$) and pyrite ($ \sim 1\times 10^{17} (\Omega m s)^{-1} @ 900 K$) (all the values are at $10^{20} cm^{-3}$ hole concentration). Even though electrical conductivity of olivine is lower, they possess almost comparable power factor with that of pyrite FeS$_2$ which implies that one need to improve the electrical conductivity of olivine to get a better TE performance. In general, we find the olivine type of Fe$_2$GeS$_4$ and Fe$_2$GeSe$_4$ to be good thermoelectric candidates along the crystallographic directions of $b$ and $c$-axes.
\par
Overall, the less dispersive bands along the high symmetry directions is responsible for the higher thermopower in olivine Fe$_2$GeS$_4$ and Fe$_2$GeSe$_4$. We find a thermopower of above $ 300 \mu V/K$ at $300 K$ and $500 K$ at a concentration of $10^{20} cm^{-3}$ in both the compounds. This higher values of thermopower together with the assumption that these materials might possess low thermal conductivity can certainly lead to good thermoelectric performance. As suggested by Tritt et. al., the minimum thermopower a material should possess to have a $ZT \sim 1$ in the considerable range of around $160-225 \mu V K^{-1}$, assuming zero lattice thermal conductivity \cite{Tritt}.
In the present case, we find the thermopower to be high in both Fe$_2$GeS$_4$ and Fe$_2$GeSe$_4$, even at the higher carrier concentration of $10^{20} cm^{-3}$ for both electrons and holes. Earlier studies on Fe$_2$GeS$_4$ show this compound to possess a high absorption coefficient ($\sim 10^5 cm^{-1}$)\cite{Yu} which imply their usage as a good photovoltaic absorber, which is almost similar to the pyrite FeS$_2$. In the present study we also found a better thermoelectric performance of Fe$_2$GeS$_4$ compared to marcasite and pyrite FeS$_2$. This allow us to state that olivine type compounds can be used as an alternate energy source of material for both thermoelectric and photovoltaic applications. We further look forward for the experiments in order to validate the proposed nature of the solar thermoelectric behaviour of the investigated systems. 

 


\section{Conclusions}
We present here the thermoelectric properties of olivine structure type of Fe$_2$GeCh$_4$ (Ch = S, Se and Te) based on the Boltzmann semi-classical transport equation using first principle calculations. 
The investigated thermoelectric properties showed Fe$_2$GeS$_4$ and Fe$_2$GeSe$_4$ to have a high thermopower above $300 \mu V/K$ even at room temperature and above, which is unusual and  can be placed in different regime of the thermoelectric family. We found a good agreement between the calculated hole concentration with the experimental measured carrier concentration of Fe$_2$GeS$_4$ at room temperature. The other interesting feature of these materials is the negligible anisotropy in the thermopower for the above two compounds, whereas electrical conductivity is two order less along $a$-axis compared with other two axes of $b$, $c$. Bipolar thermoelectric nature is observed in Fe$_2$GeTe$_4$ because of the narrow band gap. The calculated thermopower of the olivine type Fe$_2$Ge$Ch_4$ are found to be higher when compared with the marcasite and pyrite structures. Among the investigated systems Fe$_2$GeS$_4$ and Fe$_2$GeSe$_4$ are shown to have a good thermoelectric properties especially along $b$ and $c$ axes and presents substantial scope for future investigation in order to improve the TE performance.

\section{Acknowledgement}
V.K. and V.K.G would like to acknowledge IIT-Hyderabad for providing computational facility. G.V thank Center for Modelling Simulation and Design-University of Hyderabad (CMSD-UoH) for providing computational facility. V.K.G. would like to thank MHRD for the fellowship. 

\clearpage
\newpage


\begin{thebibliography}{0000}

\bibitem{Furdyna} Furdyna J K, Kossut J, (ed) Willardson R K and Beer A C, in: Diluted Magnetic Semiconductors, in Semiconductors and Semimetals, 1988, Vol. \textbf{25}, Academic Press, New York 1988.


\bibitem{Tore1} Ericsson T, Holényi K and Amcoff O 1997 \textit{J. Phys.: Condens. Matter} \textbf{9} 3943.  

\bibitem{Tore2} Baron V, Amcoff O and Ericsson T 1999 \textit{J. Magnetism and Magnetic Materials} \textbf{195} 81-92.  

\bibitem{Quintero} Quintero M, Ferrer D, Caldera D, Moreno E, Quintero E, Morocoima M, Grima P, Bocaranda P, Delgado G E and Henao J A 2009 \textit{Journal of Alloys and Compounds} \textbf{469} 4-8.	

\bibitem{Junod} Junod A, Wang K -O, Triscone G and Larnarche G 1995 \textit{Journal of Magnetism and Magnetic Materials} \textbf{146} 21-29. 


\bibitem{Guo} Jiang X, Guo G Y 2004 \textit{Phys. Rev. B} \textbf{69} 155108.	

\bibitem{Yu} Yu L, Lany S, Kykyneshi R, Jieratum V, Ravichandran R, Pelatt B, Altschul E, Platt H A S, Wager J F Keszler D A,  Zunger A 2011 \textit{Adv. Energy Mater.} \textbf{1} 748-753.	

\bibitem{Fredrick} Fredrick S J and Prieto A L 2013 \textit{J. Am. Chem. Soc.} \textbf{135} 18256-18259.	

\bibitem{Park} Park B -I, Yu S, Hwang Y, Cho S -H, Lee J -S, Park C, Lee D -K and Lee S Y 2015 \textit{J. Mater. Chem. A} \textbf{3} 2265-2270.	

\bibitem{thesis} Platt H A S 2010 \textit{Copper and Iron Chalcogenides for Efficient Solar Absorption (Oregon State University)} (Thesis).	

\bibitem{TAGS} Hsu K F, Loo S, Guo F, Chen W, Dyck J S, Uher C, Hogan T, Polychroniadis E K and Kanatzidis M G 2004 \textit{Science} \textbf{303} 818.	

\bibitem{Sales}Sales B C, Mandrus D and Williams R K 1996 \textit{Science} \textbf{272} 1325.	

\bibitem{YPei} Pei Y, Shi X, LaLonde A, Wang H, Chen L and Snyder G J 2011 \textit{Nature} \textbf{473} 66.	

\bibitem{Blaha} Blaha P, Schwarz K, Madsen G K H, Kvasnicka D and Luitz J 2001 WIEN2K, An Augmented Plane Wave + Local Orbitals Program for Calculating Crystal Properties (Karlheinz Schwarz, Techn. Universit\"{a}t Wien, Austria) $<http://www.wien2k.at/>$.	

\bibitem{Becke} Becke A D and Johnson E R 2006 {\it J. Chem. Phys.} {\bf 124} 221101-1-4. 

\bibitem{Tran1} Tran F and Blaha P 2009 {\it Phys. Rev. Lett.} {\bf 102} 226401-1-4. 

\bibitem{Perdew} Perdew J P, Burke K and Ernzerhof M 1996 {\it Phys. Rev. Lett.} {\bf 77} 3865-3868. 

\bibitem{Monkhorst} Monkhorst H J and Pack J D 1976 {\it Phys. Rev. B} {\bf 13} 5188-5192. 

\bibitem {Madsen} Madsen G K H and Singh D J 2006 {\it Comput. Phys. Commun.} {\bf 175} 67−71.	

\bibitem{Scheidemantel} Scheidemantel T J, Ambrosch-Draxl C, Thonhauser T, Badding J V and Sofo J O 2003 {\it Phys. Rev. B} {\bf 68} 125210-1-6. 

\bibitem{Jodin} Jodin L, Tobola J, P\'{e}cheur P, Scherrer H and Kaprzyk S 2004 {\it Phys. Rev. B} {\bf 70} 184207-1-11. 

\bibitem{Chaput} Chaput L, P$\acute{e}$cheur P, Tobola J and Scherrer H 2005 {\it Phys. Rev. B} {\bf 72} 085126-1-11. 

\bibitem{Khuong} Ong K P, Singh D J and Wu P 2011 {\it Phys. Rev. B} {\bf 83} 115110-1-5. 

\bibitem{singh} Singh D J 2010 {\it Func. Mat. Lett.} {\bf 3} 223-226. 

\bibitem{aggate2} Parker D and Singh D J 2012  {\it Phys. Rev. B} {\bf 85} 125209-1-7. 

\bibitem{FeX2} Gudelli V K, Kanchana V, Vaitheeswaran G, Valsakumar M C and Mahanti S D 2014 \textit{RSC Adv.} \textbf{4} 9424-9431.	

\bibitem{vesta}Momma K and Izumi F 2011 \textit{J. Appl. Crystallogr.} \textbf{44} 1272–1276. 

\bibitem{xcrysden} Kokalj A 2003 \textit{Comput. Mater. Sci.} \textbf{28} 155.	

\bibitem{FeS2} Gudelli V K, Kanchana V, Appalakondaiah S, Vaitheeswaran G and Valsakumar M C 2013 \textit{J. Phys. Chem. C} \textbf{117} 21120-21131.	

\bibitem{Fe2GeS4}Vincent H, Bertuat E F, Baur W H and Shannon R D 1976 \textit{Acta Cryst. B} \textbf{32} 1749-1755. 

\bibitem{Fe2GeSe4} Henao J A, Delgado J M, Quintero M 1998 \textit{Powder Diffraction} \textbf{13(4)} 196-201.

\bibitem{Fe2GeTe4} Delgado G E, Betancourt L, Mora A J, Contreras J E, Grima-Gallardo P and Quintero M 2010 \textit{Chalcogenide Letters} \textbf{7} 133 - 138.

\bibitem{Dixit} Dixit H, Saniz R, Cottenier S, Lamoen D and Partoens B 2012 {\it J. Phys.: Condens. Matter} {\bf 24} 205503-1-9.	

\bibitem{Nagaraja} Nagaraja A R, Perry N H, Mason T O, Tang Y, Grayson M, Paudel T R, Lany S and Zunger A 2011 \textit{J. Am. Ceram. Soc.} \textbf{95} 1–6. 

\bibitem{Volnianska} Volnianska O and Boguslawski P 2013 \textit{J. Appl. Phys.} \textbf{114} 033711.	

\bibitem{Rosso} Rosso K M Becker U and Hochella Jr M F 1999 \textit{Amer. Miner.} \textbf{84} 1535-1548.	

\bibitem{Eyret} Eyert V, H\"{o}ck K -H, Fiechter S and Tributsch H 1998 {\it Phys. Rev. B} {\bf  57}, {\it}, 6350-6359.	

\bibitem{SnSe} Zhao L -D, Lo S -H, Zhang Y, Sun H, Tan G, Uher C, Wolverton C, Dravid V P and Kanatzidis M G 2014 \textit{Nature} \textbf{508} 373–377.	

\bibitem{DJS_PRL} Ong K P Singh D J and Wu P 2010 \textit{Phy. Rev. Lett.} \textbf{104} 176601.	

\bibitem{JAP_CuGaTe2} Gudelli V K, Kanchana V, Vaitheeswaran G, Svane A and Christensen N E 2013 \textit{J. Appl. Phys.} \textbf{114} 223707.	

\bibitem{Stordeur} Stordeur M and K€hnberger W 1975 \textit{Phys. Status Solidi B} \textbf{69} 377-387.	

\bibitem{Testardi}Testardi L R, Bierly Jr J N and Donahoe F J 1962 \textit{J. Phys. Chem. Solids} \textbf{23} 1209-1217.	

\bibitem{Jeon} Jeon H -W, Ha H -P, Hyun D -B and Shim J -D 1991 \textit{J. Phys. Chem. Solids} \textbf{52} 579–585. 

\bibitem{Plechcek}Plechcek T, Navrtil J, Hork J and Lot’k P 2004 \textit{Philos. Mag.} \textbf{84} 2217–2228.	

\bibitem{Tritt} Tritt T M and Subramanian M A 2006 \textit{MRS Bulletin} \textbf{31} 188.	

\end{thebibliography}

\clearpage
\newpage

\begin{table}
\caption{The calculated effective mass of Fe$_2$GeCh$_4$ ($Ch$ = S, Se, Te) in crystallographic directions of the Brillouin zone in the units of electron rest mass}
\begin{center}
\begin{tabular}{cccccccccccccccccccccc}
\hline
Direction	&Fe$_2$GeS$_4$	&Fe$_2$GeSe$_4$	&Fe$_2$GeTe$_4$\\
\hline
VBM\\
\hline
$\Gamma$-X		&6.80		&3.11		&0.29	\\			
$\Gamma$-Y		&4.16		&3.91		&0.04	\\
$\Gamma$-Z		&4.21		&4.18		&0.53	\\
\hline
CBM\\	
\hline
$\Gamma$-X		&3.37		&3.03		&0.98	\\
$\Gamma$-Y		&3.39		&3.02		&0.91	\\
$\Gamma$-Z		&1.60		&2.15		&2.24	\\
\hline
\end{tabular}
\end{center}
\end{table}

\begin{figure*}
\begin{center}
\subfigure[]{\includegraphics[width=80mm,height=60mm]{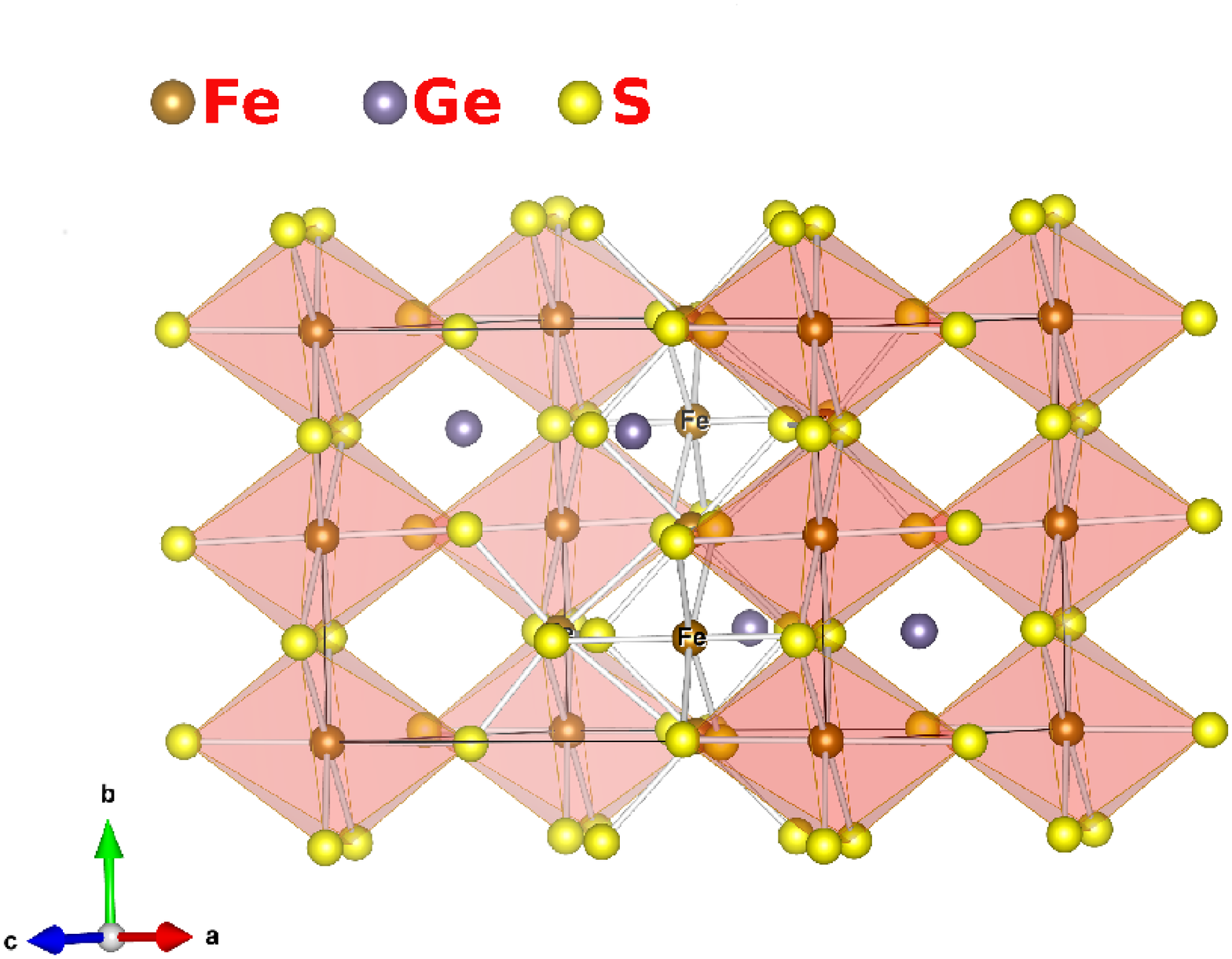}}
\subfigure[]{\includegraphics[width=80mm,height=50mm]{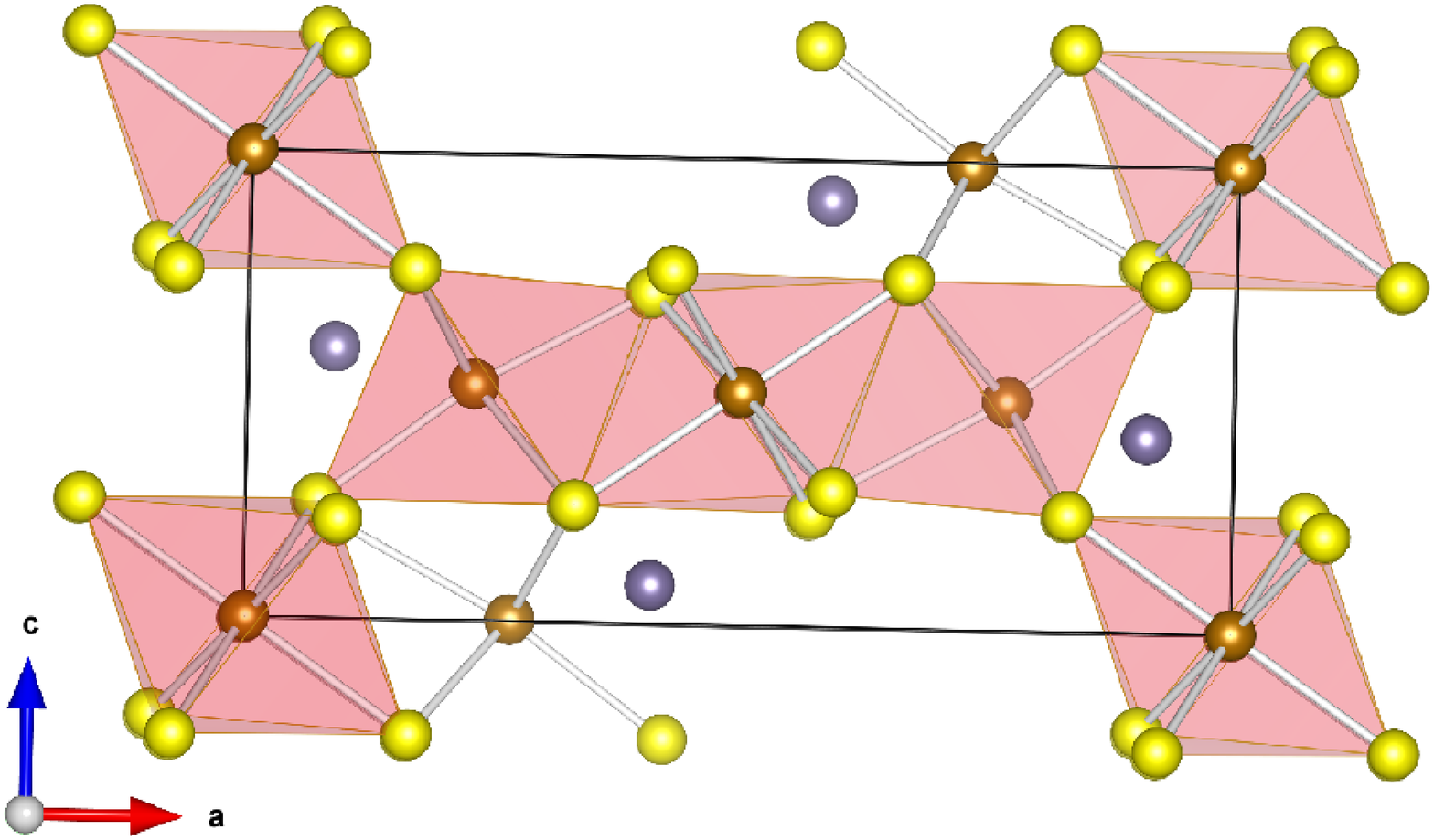}}\\
\subfigure[]{\includegraphics[width=65mm,height=65mm]{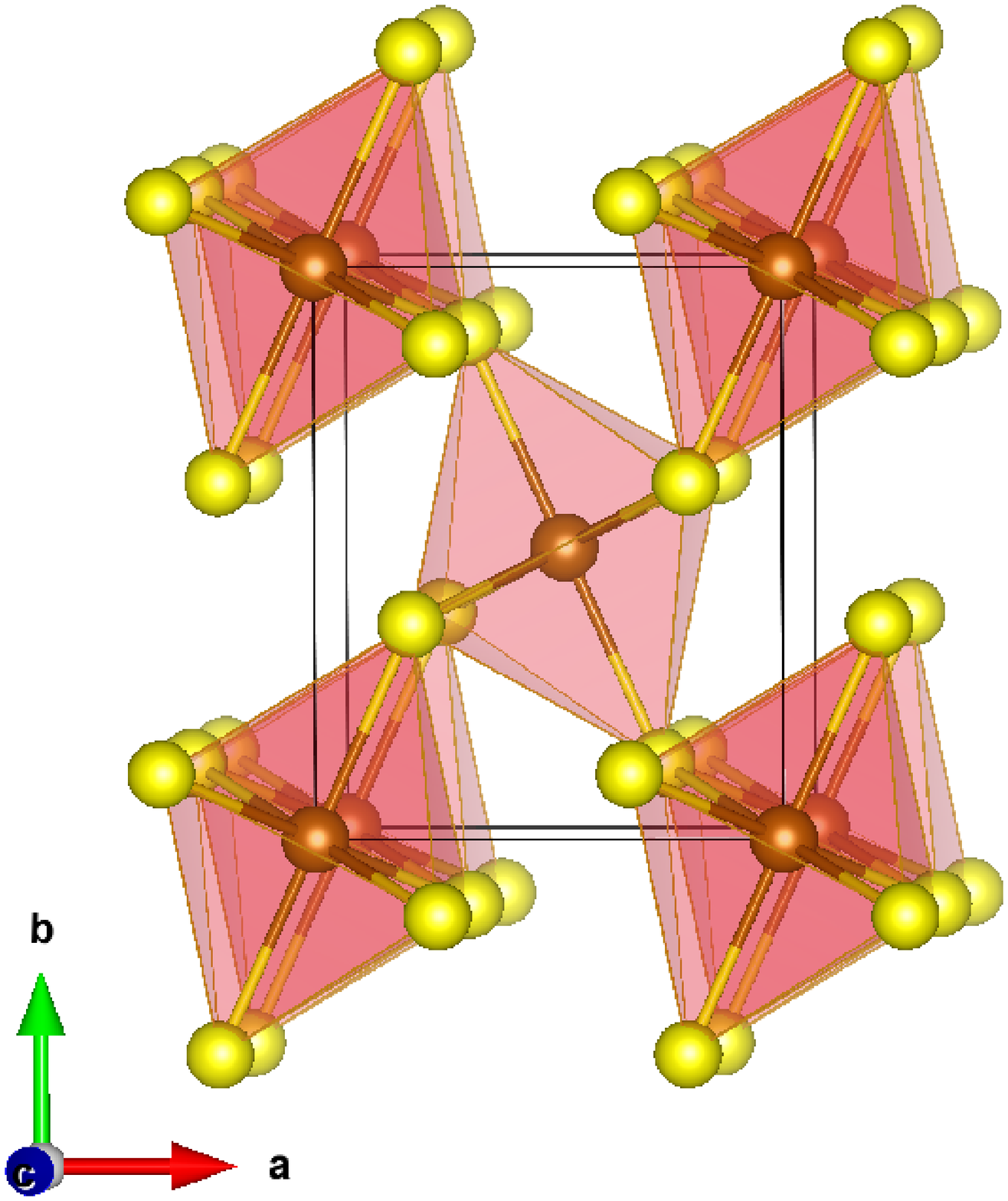}}
\caption{(Color online) Crystal structure of (a) edge-sharing octahedron (b) vertex-sharing octahedron of Fe$_2$GeS$_4$  compared with (c) marcasite FeS$_2$}
\end{center}
\end{figure*}

\begin{figure*}
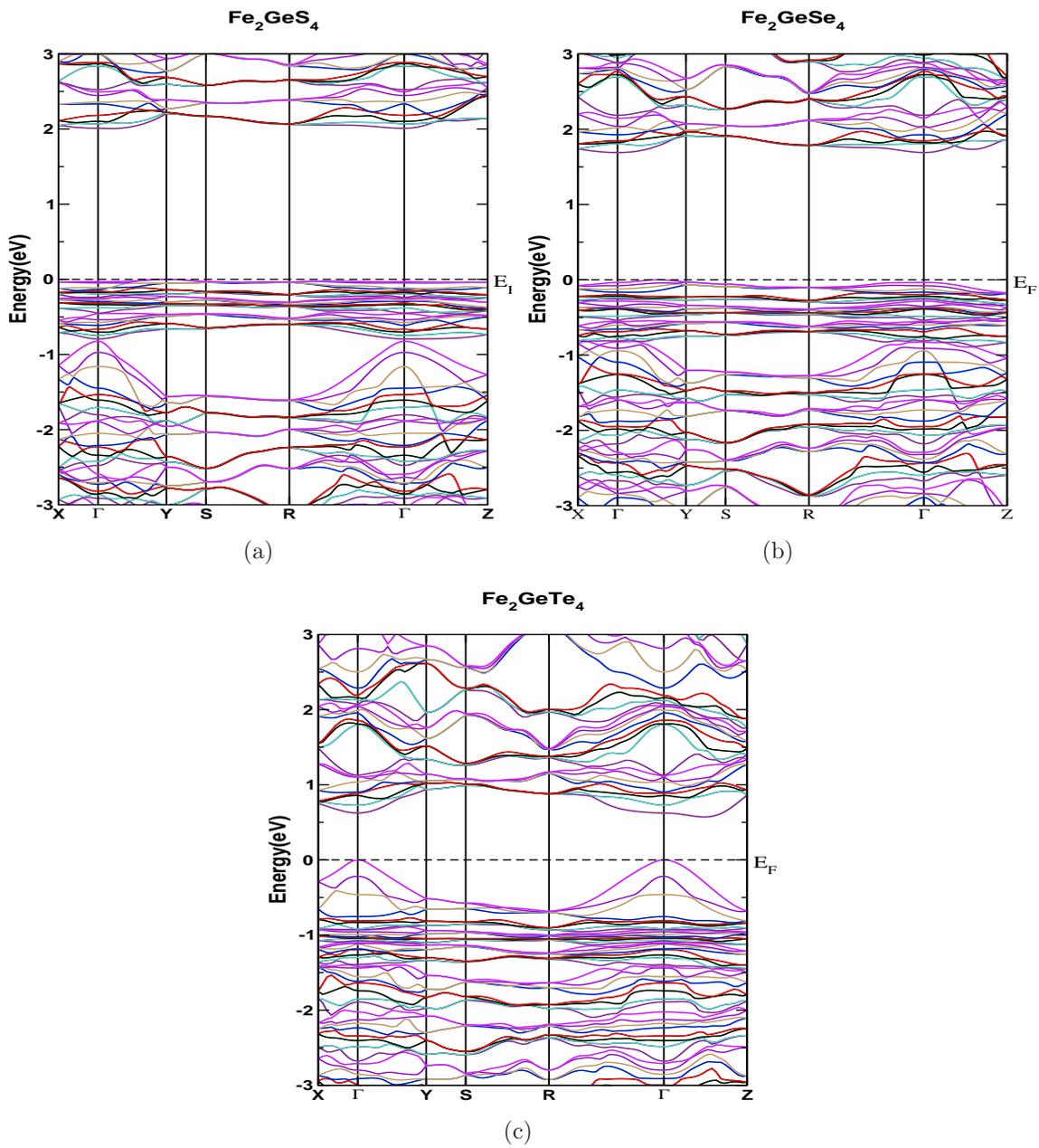

\begin{center}
\subfigure[]{\includegraphics[width=75mm,height=75mm]{4.eps}}
\subfigure[]{\includegraphics[width=75mm,height=75mm]{5.eps}}\\
\subfigure[]{\includegraphics[width=75mm,height=75mm]{6.eps}}
\caption{(Color online) Calculated band structure of (a) Fe$_2$GeS$_4$ (b) Fe$_2$GeSe$_4$ and (c) Fe$_2$GeTe$_4$}
\end{center}
\end{figure*}

\begin{figure*}
\begin{center}
\includegraphics[width=60mm,height=50mm]{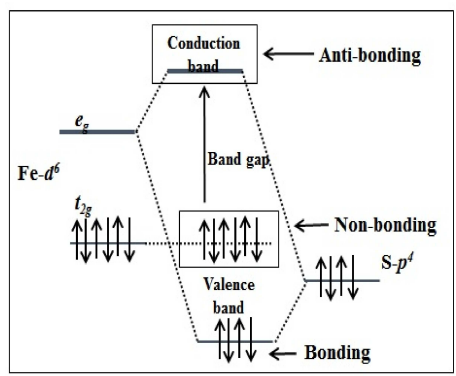}
\caption{(Color online) Schematic representation of octahedronl crystal field splitting of $Fe-d$ states in Fe$_2$GeCh$_4$}
\end{center}
\end{figure*}

\begin{figure*}
\begin{center}
\subfigure[]{\includegraphics[width=160mm,height=70mm]{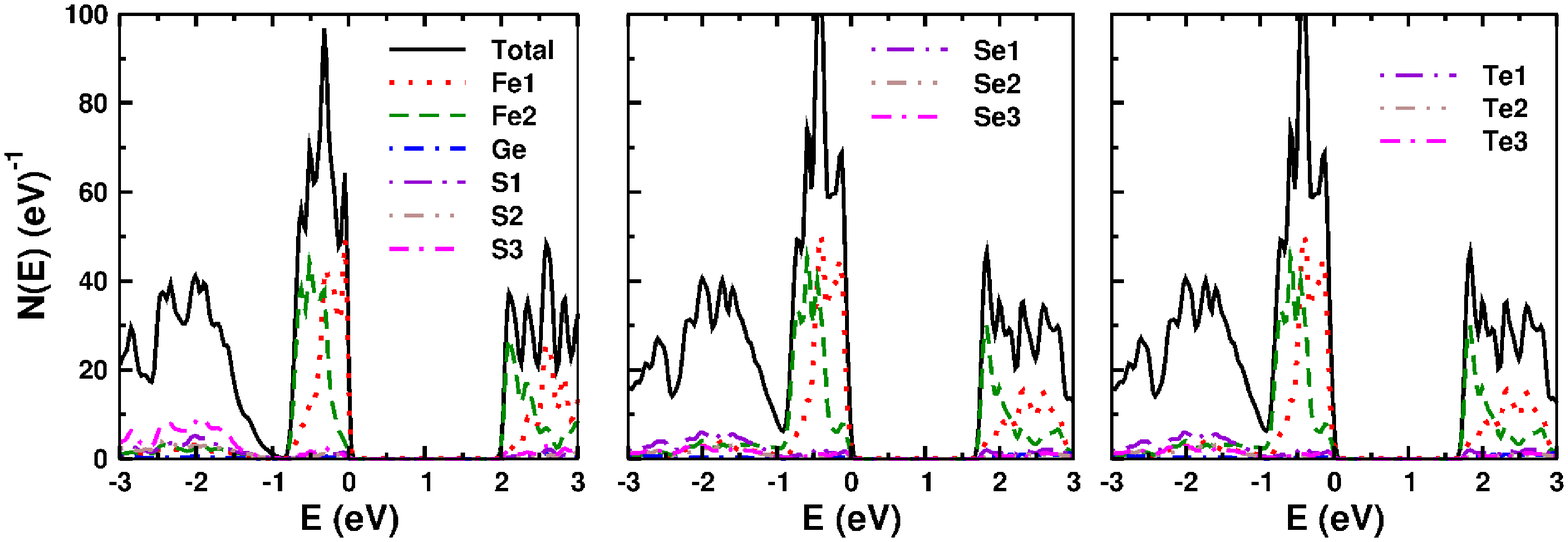}}\\
\subfigure[]{\includegraphics[width=75mm,height=70mm]{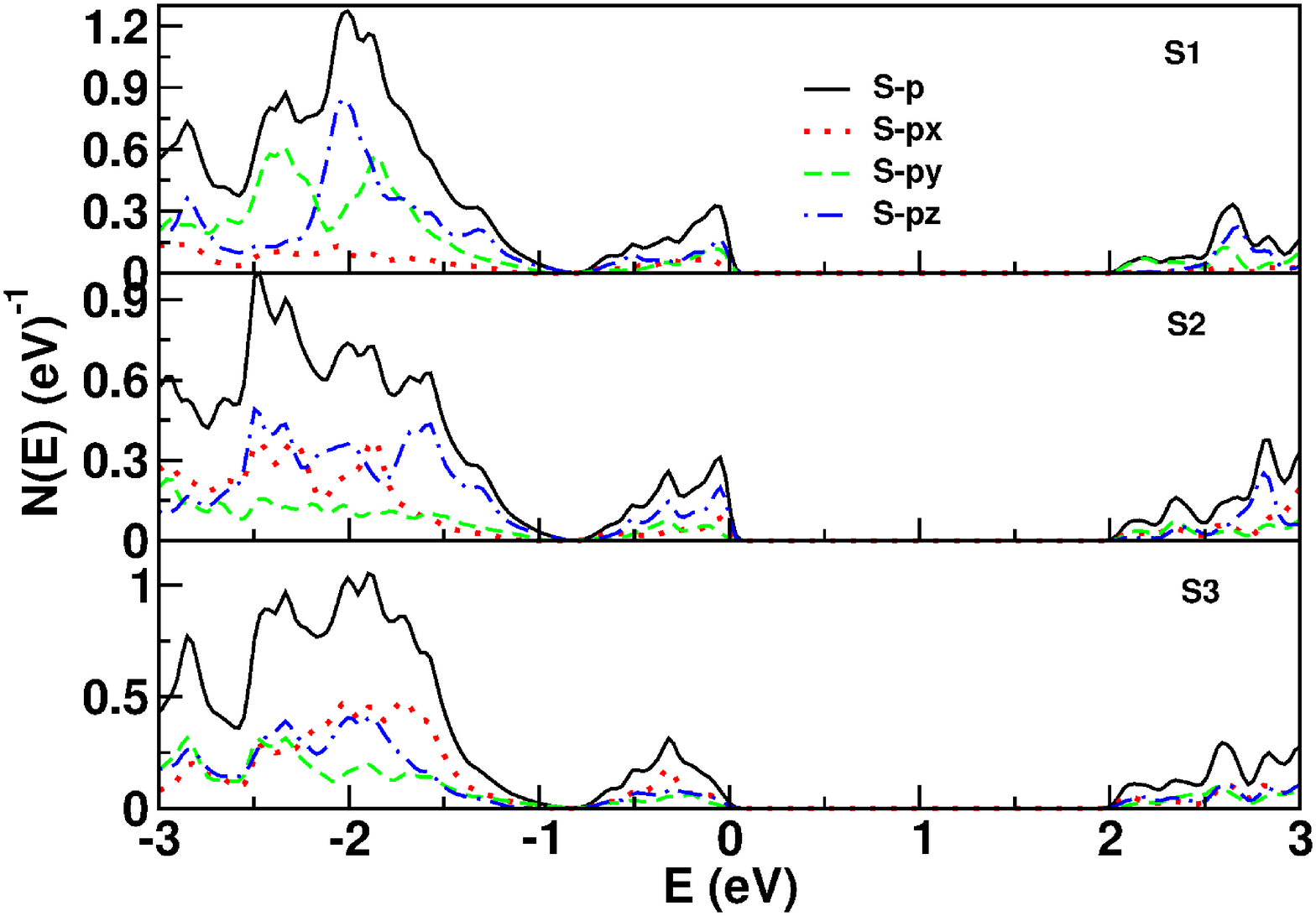}}\\
\subfigure[]{\includegraphics[width=75mm,height=70mm]{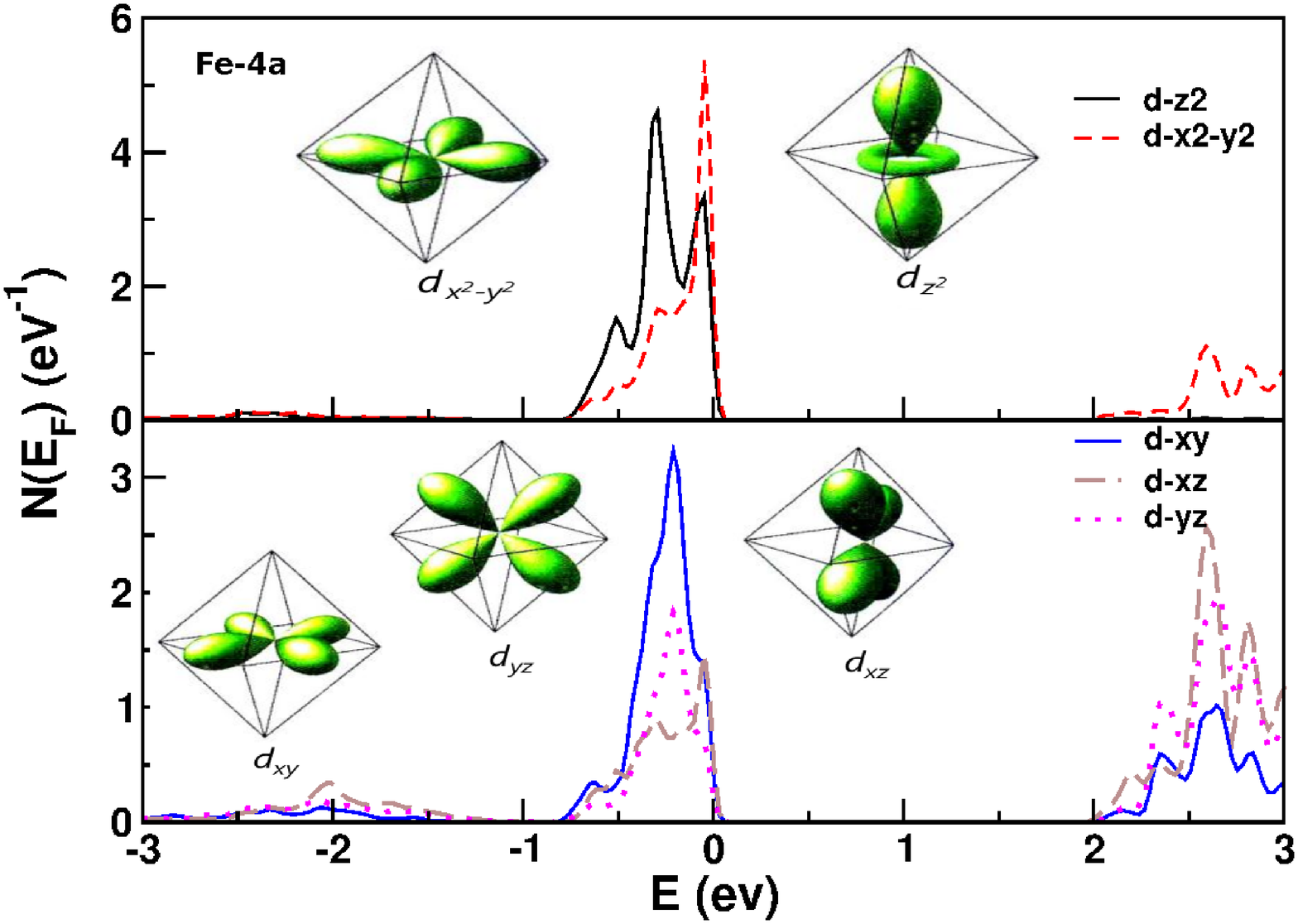}}
\subfigure[]{\includegraphics[width=75mm,height=70mm]{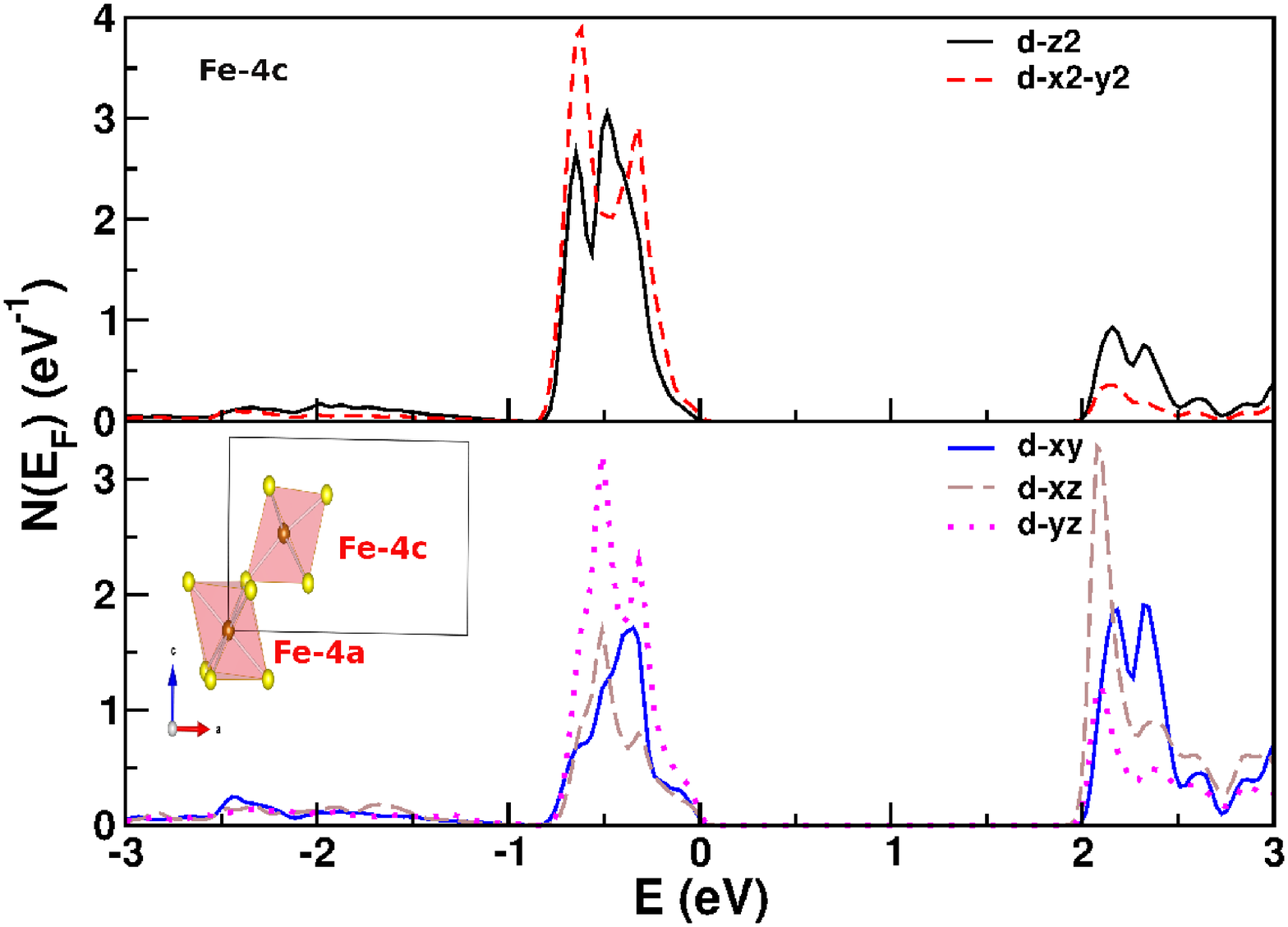}}\\
\caption{(Color online) (a) Calculated DOS of Fe$_2$GeCh$_4$, and m-projected DOS for (b) S-p, (c) Fe1 at 4a and (d) Fe2 at 4c positions of Fe$_2$GeS$_4$.}
\end{center}
\end{figure*}

\begin{figure*}
\begin{center}
\subfigure[]{\includegraphics[width=55mm,height=50mm]{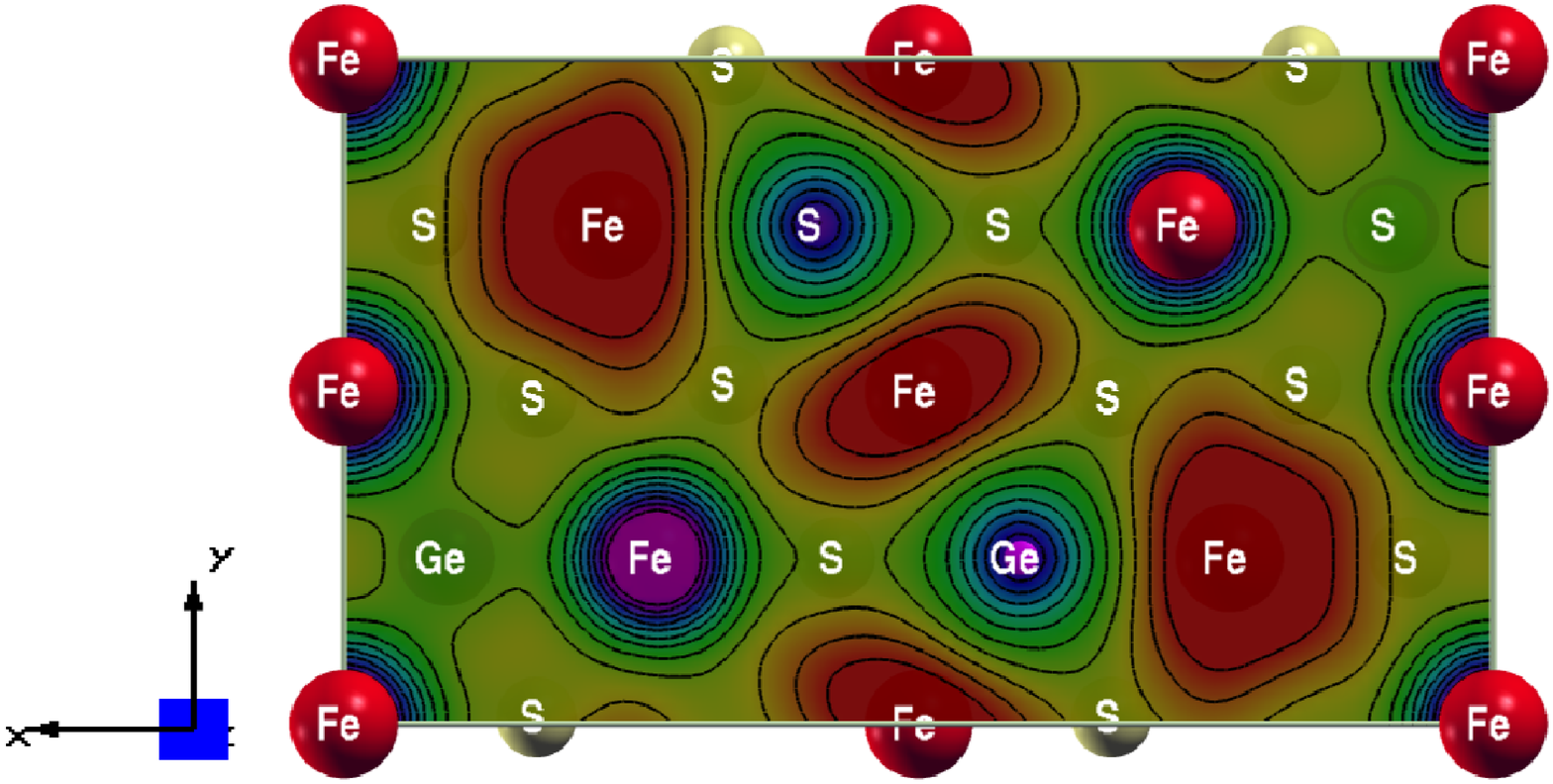}}
\subfigure[]{\includegraphics[width=55mm,height=50mm]{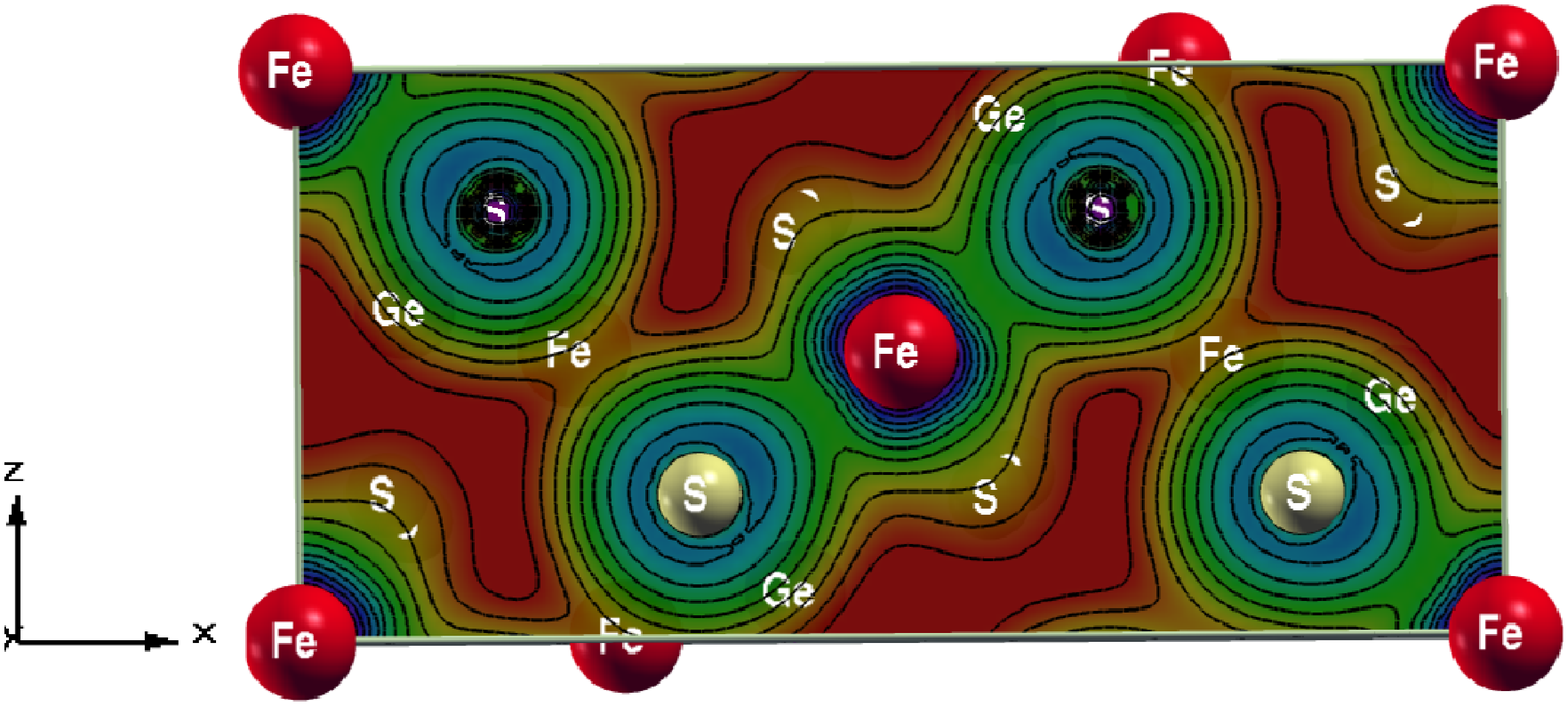}}
\subfigure[]{\includegraphics[width=50mm,height=50mm]{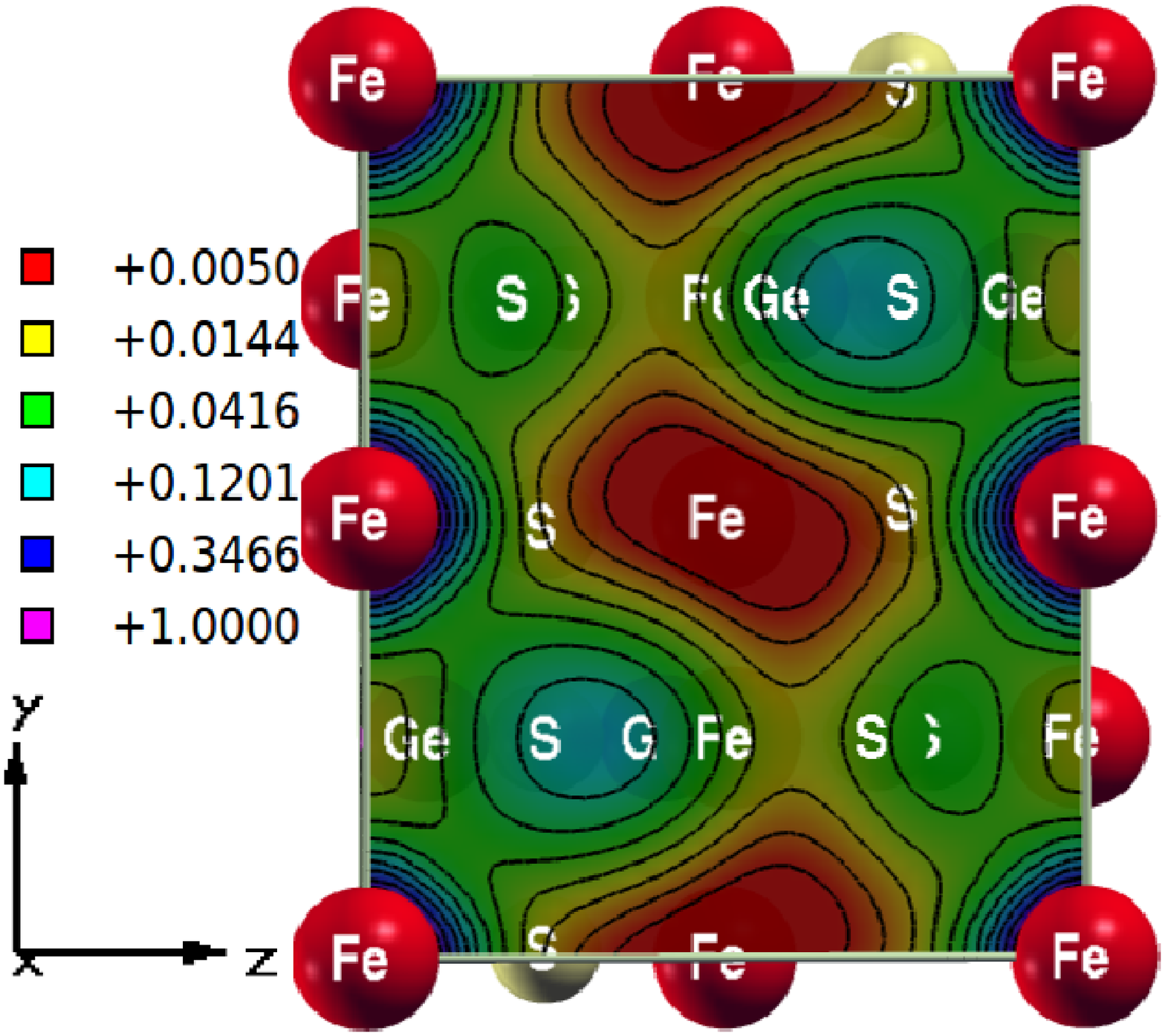}}
\caption{(Color online) Calculated charge density density along (a) $xy$ (b) $xz$ and (c) $yz$ planes of Fe$_2$GeS$_4$}
\end{center}
\end{figure*}

\begin{figure*}
\begin{center}
\subfigure[]{\includegraphics[width=120mm,height=60mm]{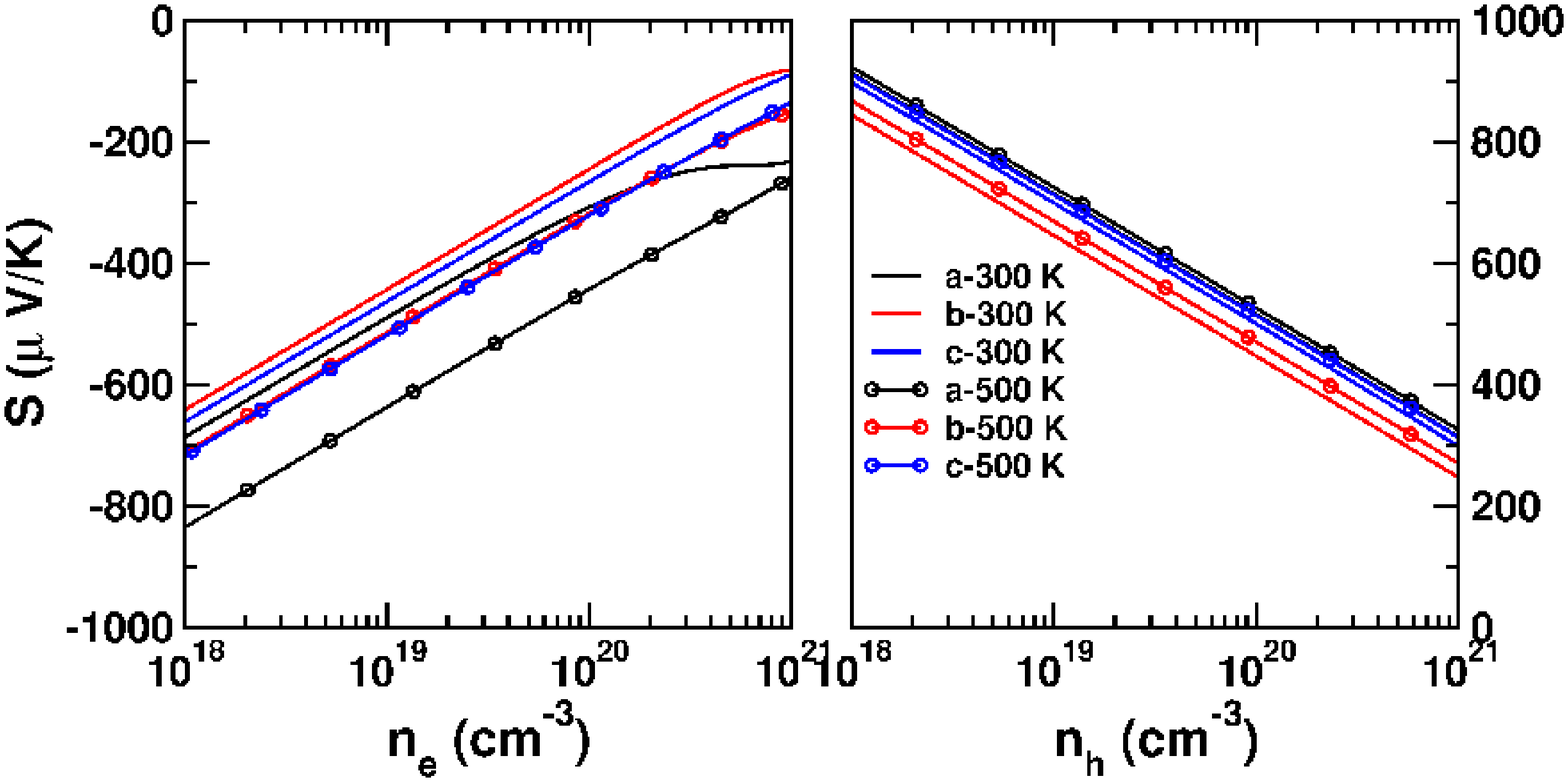}}\\
\subfigure[]{\includegraphics[width=120mm,height=60mm]{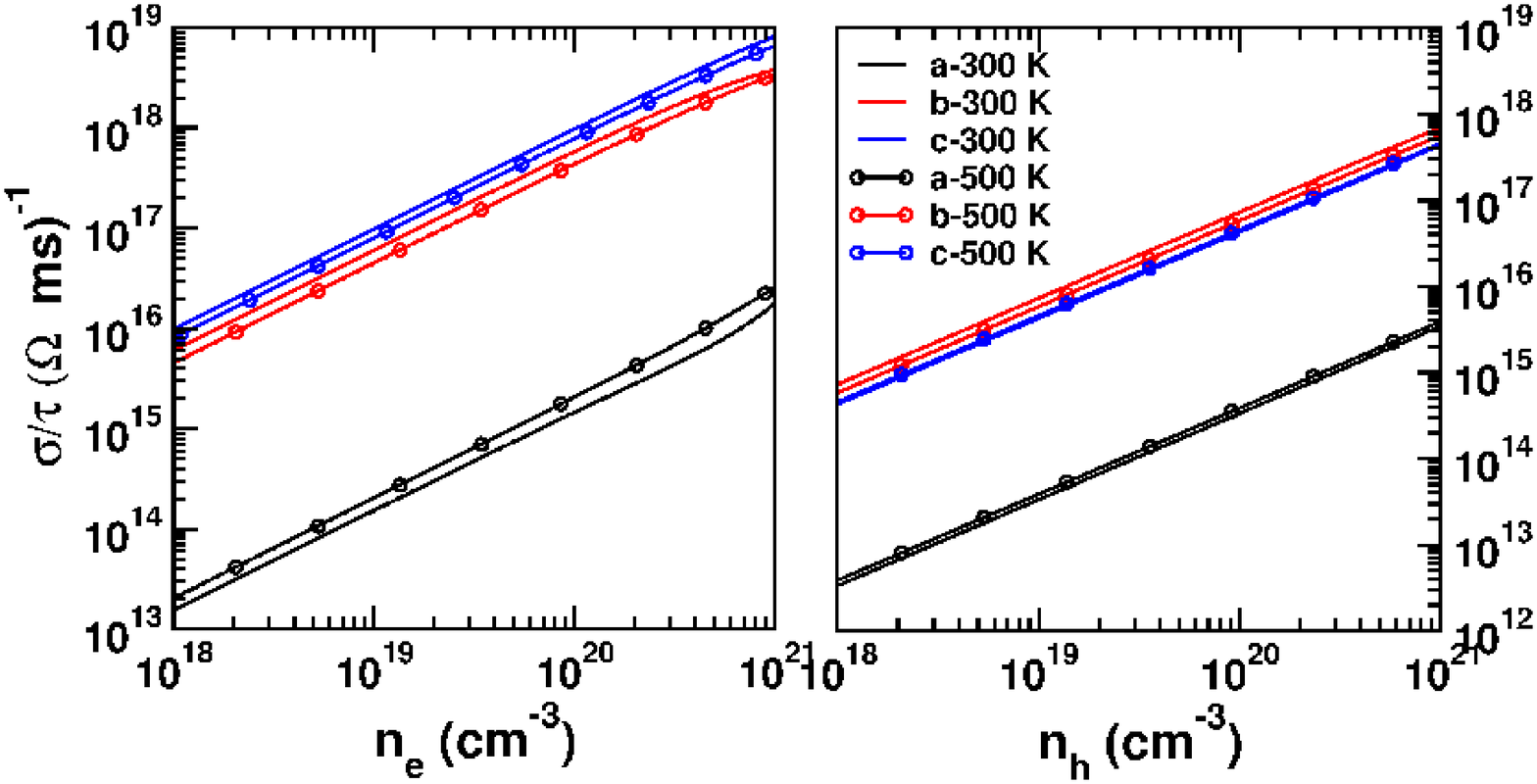}}\\
\subfigure[]{\includegraphics[width=120mm,height=60mm]{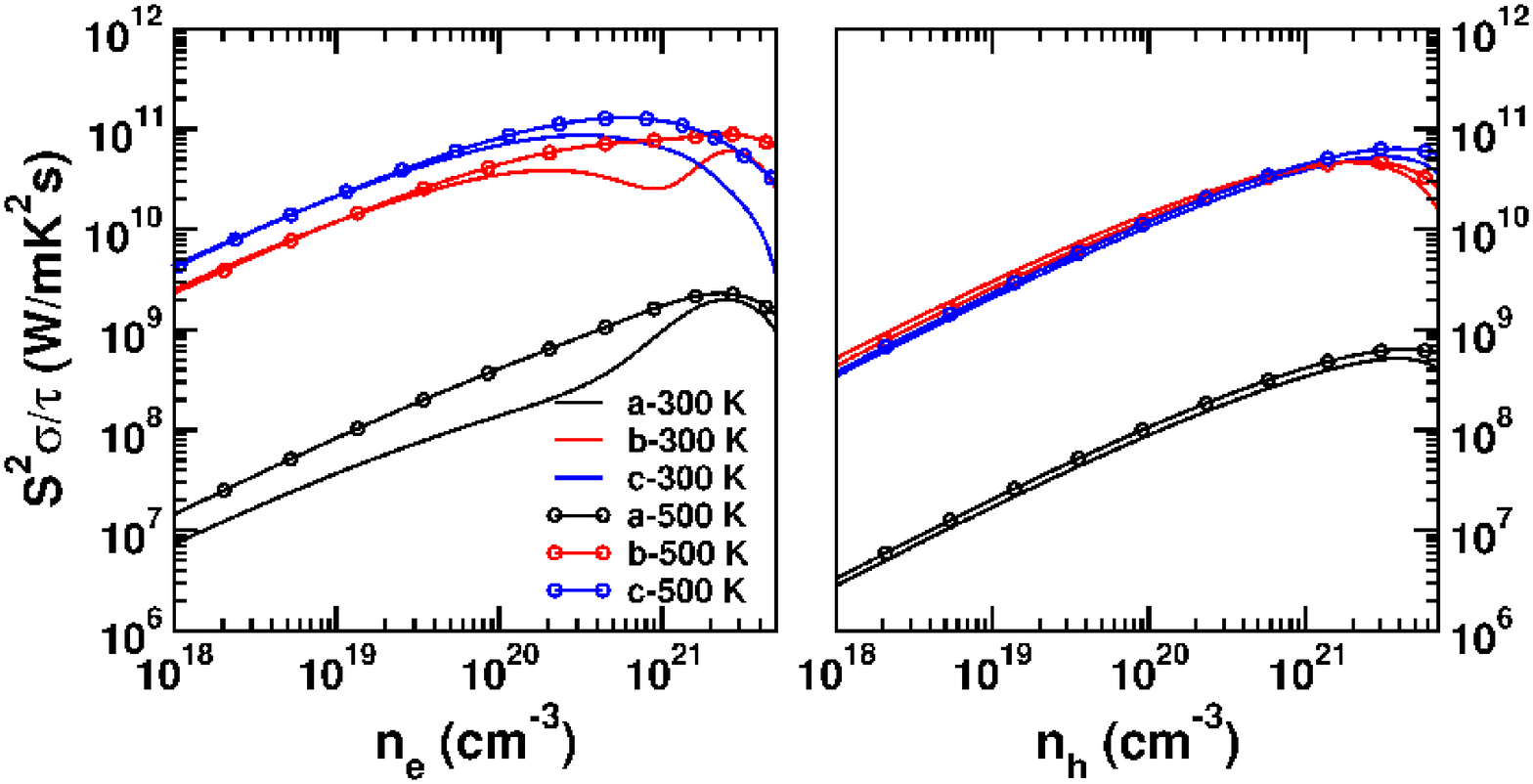}}\\
\caption{(Color online) Calculated (a) thermopower (b) electrical conductivity scaled by relaxation time and (c) Power factor of Fe$_2$GeS$_4$}
\end{center}
\end{figure*}

\begin{figure*}
\begin{center}
\subfigure[]{\includegraphics[width=120mm,height=60mm]{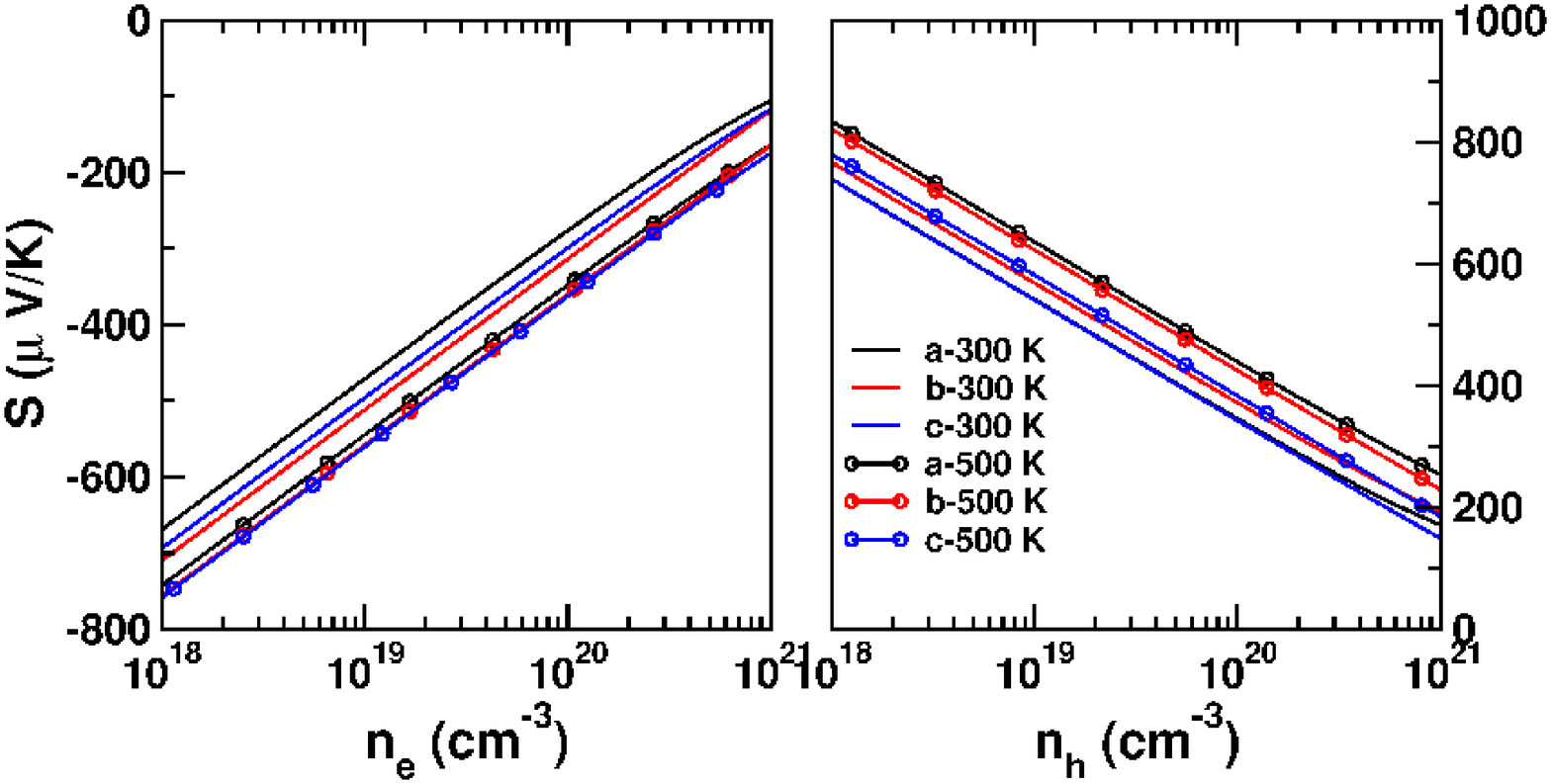}}\\
\subfigure[]{\includegraphics[width=120mm,height=60mm]{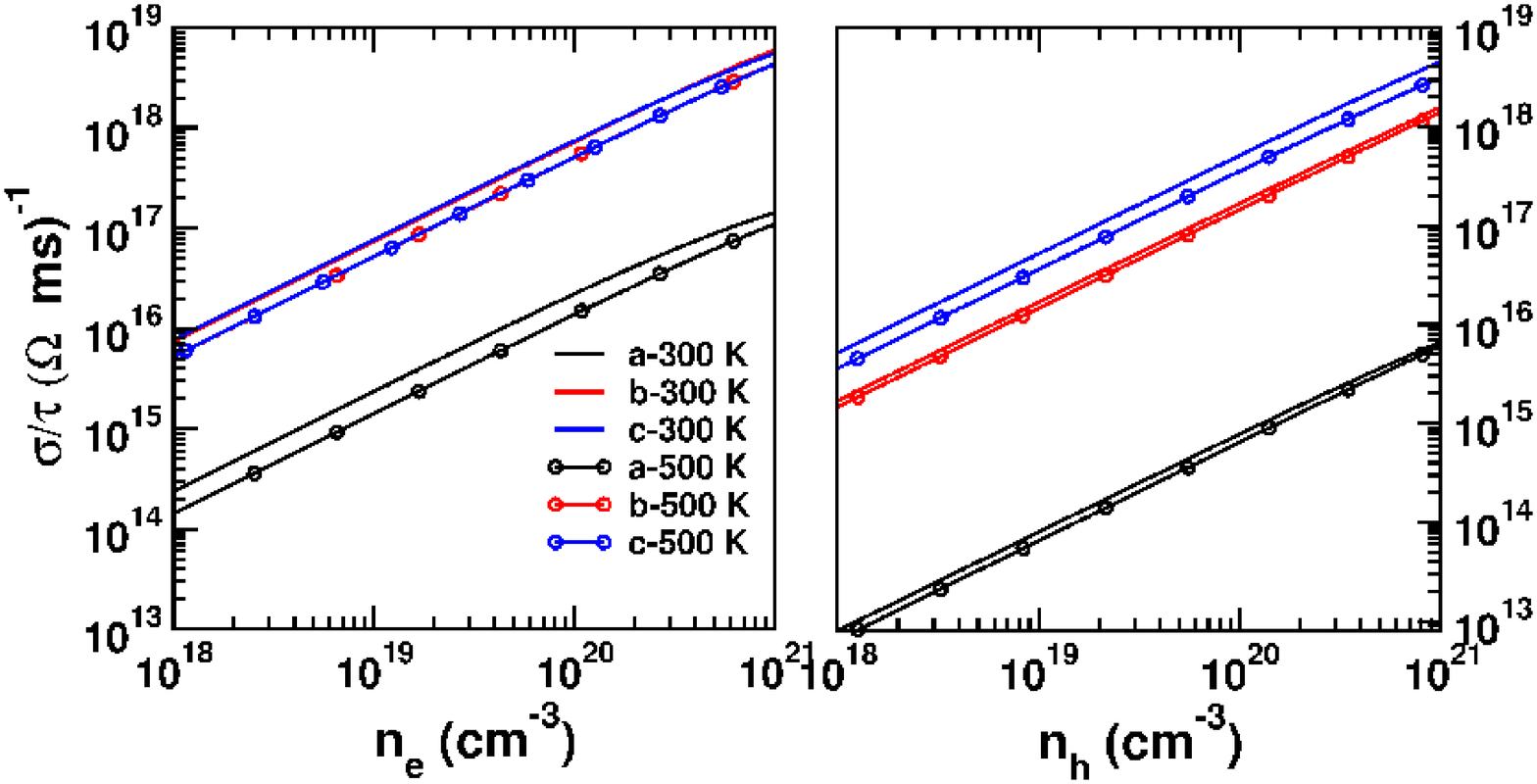}}\\
\subfigure[]{\includegraphics[width=120mm,height=60mm]{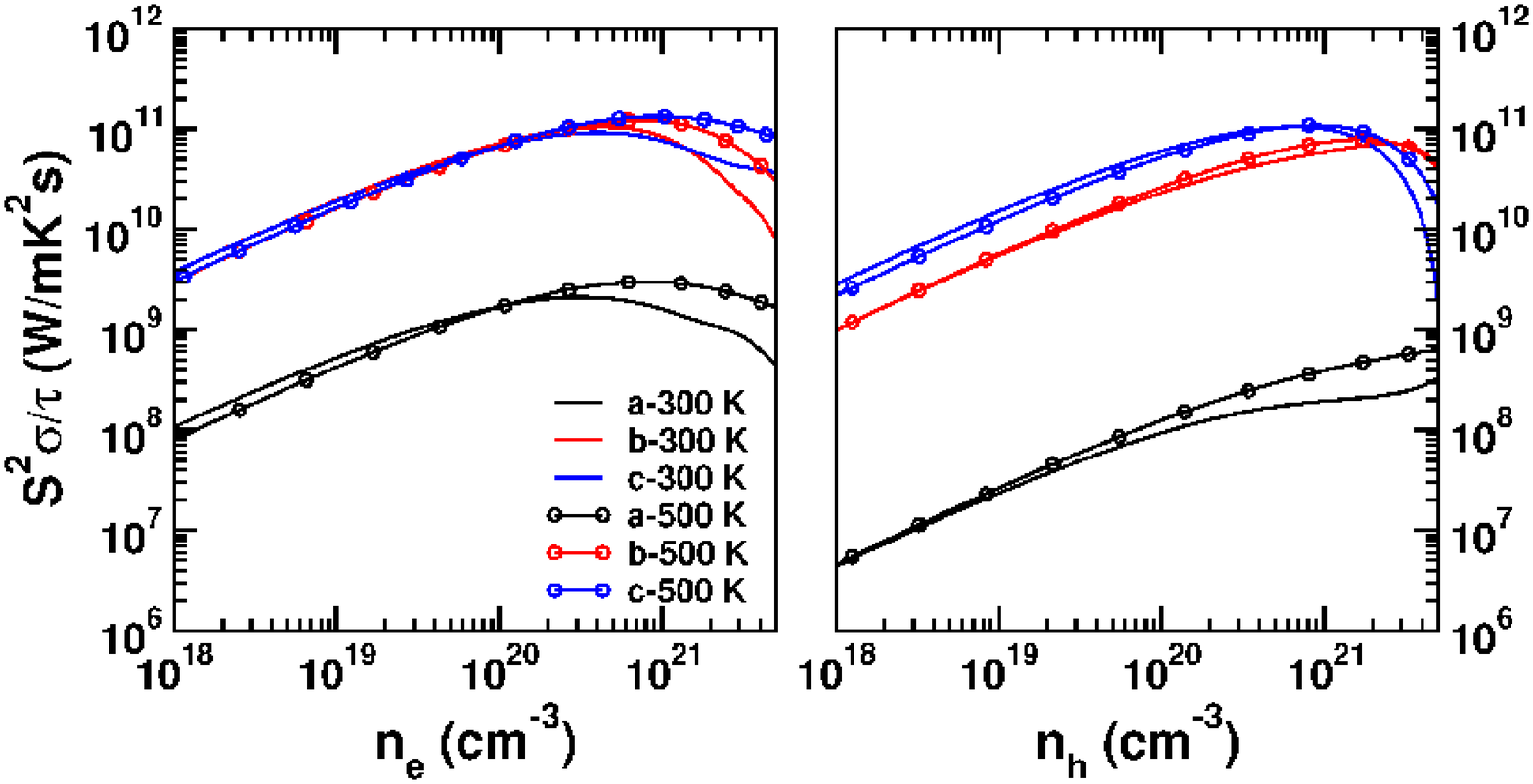}}\\
\caption{(Color online) Calculated (a) thermopower (b) electrical conductivity scaled by relaxation time and (c) Power factor of Fe$_2$GeSe$_4$}
\end{center}
\end{figure*}

\begin{figure*}
\begin{center}
\subfigure[]{\includegraphics[width=120mm,height=60mm]{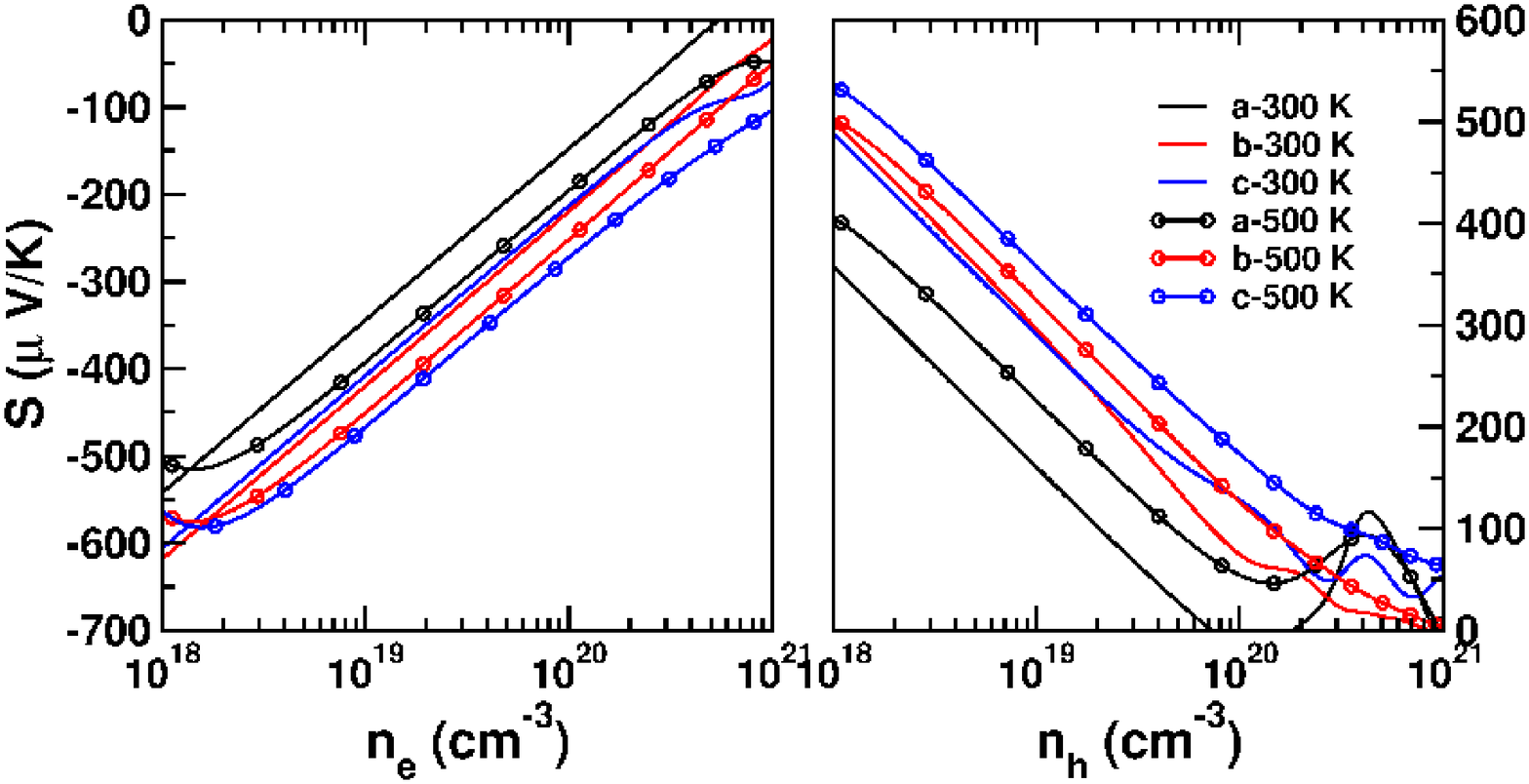}}\\
\subfigure[]{\includegraphics[width=120mm,height=60mm]{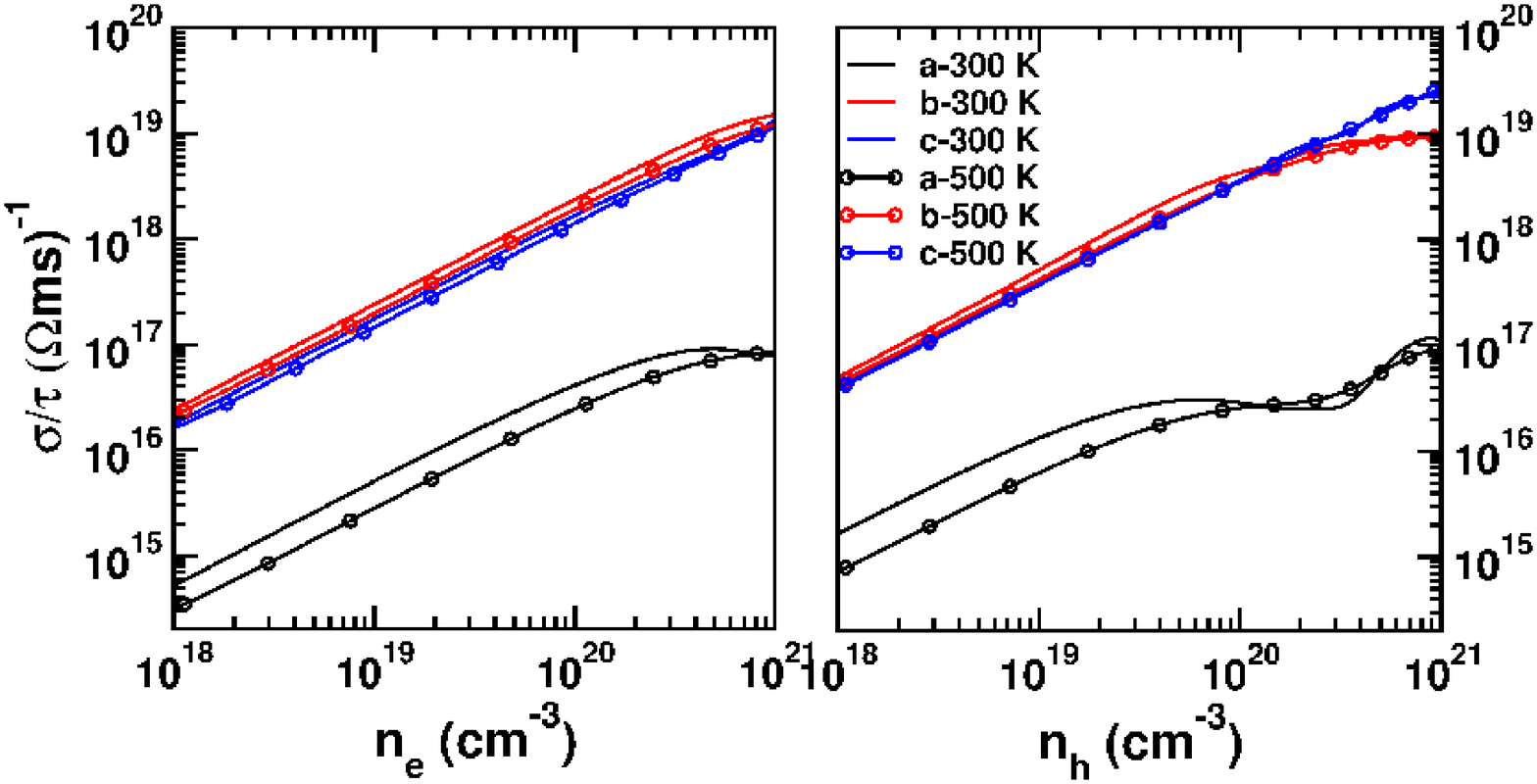}}\\
\caption{(Color online) Calculated (a) thermopower (b) electrical conductivity scaled by relaxation time of Fe$_2$GeTe$_4$}
\end{center}
\end{figure*}

\begin{figure*}
\begin{center}
\includegraphics[width=120mm,height=55mm]{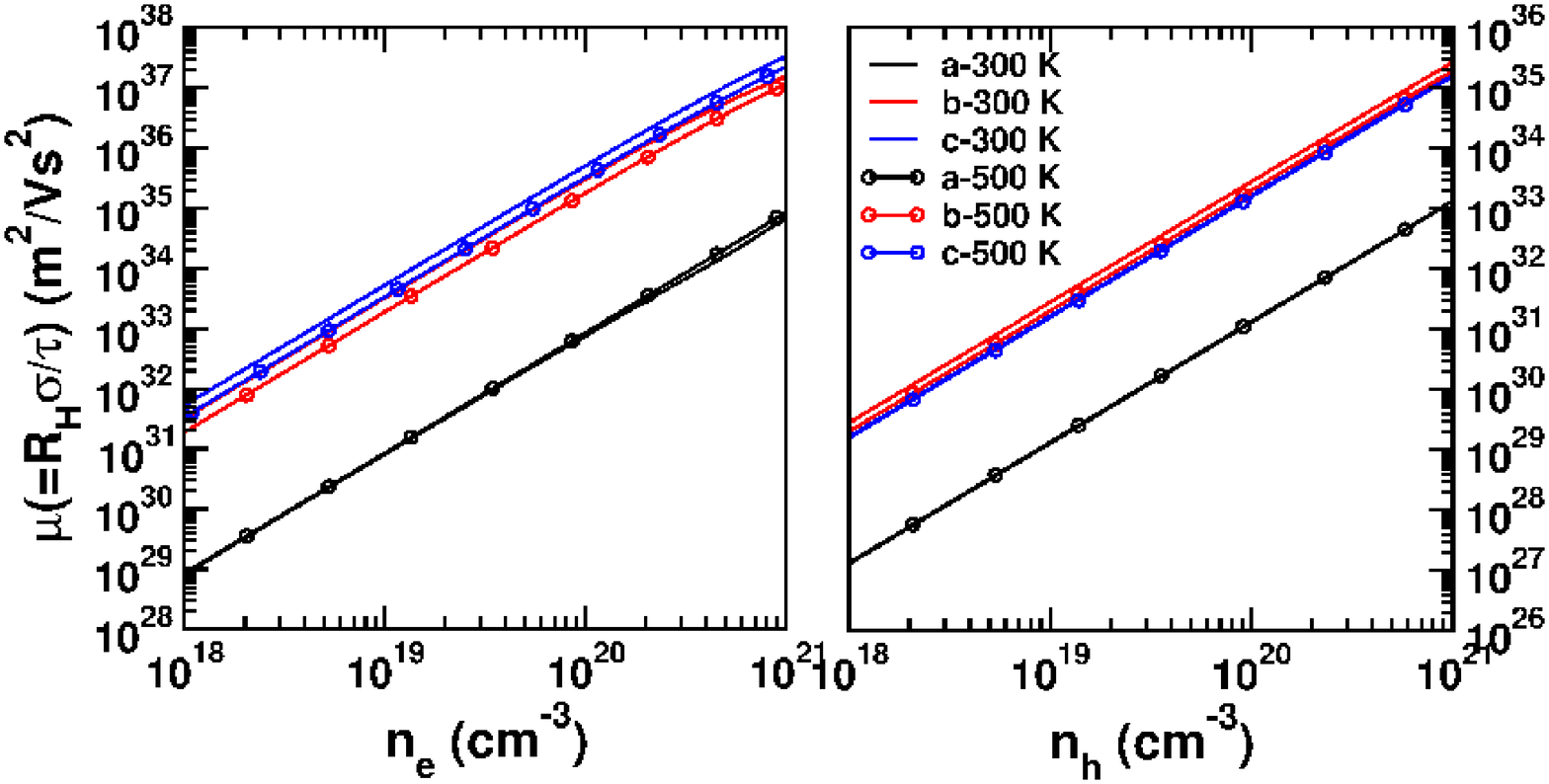}
\caption{(Color online) The Calculated mobility scaled by relaxation time of Fe$_2$GeS$_4$}
\end{center}
\end{figure*}

\end{document}